\documentclass[draftclsnofoot,twoside,onecolumn,letter,12pt]{IEEEtran}
\usepackage{graphicx}
\usepackage{amssymb}
\usepackage{amsmath}
\usepackage{cite}

\newtheorem{definition}{Definition}
\newtheorem{theorem}{Theorem}
\newtheorem{lemma}{Lemma}

\DeclareMathOperator{\tr}{\mathrm{tr}}
\DeclareMathOperator{\E}{\mathbb{E}}

\DeclareMathOperator{\sgn}{\mathrm{sgn}}

\newcommand{\EX}[1]{\E\left\{{#1}\right\}}

\newcommand{\PDF}[2]{p_{{#1}}\left({#2}\right)}

\newcommand{\CF}[2]{\Phi_{#1}\left({#2}\right)}
\newcommand{\CGF}[2]{\mathcal{K}_{#1}\left({#2}\right)}

\newcommand{\B}[1]{{\mathbf{#1}}}
\newcommand{\C}{\mathbb{C}}

\newcommand{\CGM}[5]{\tilde{\mathcal{N}}_{{#1},{#2}}\left({#3},{#4},{#5}\right)}
\newcommand{\CWM}[3]{\tilde{\mathcal{W}}_{#1}\left({#2},{#3}\right)}
\newcommand{\CQM}[5]{\tilde{Q}_{{#1},{#2}}\left({#3},{#4},{#5}\right)}
\newcommand{\nTx}{{n_{\mathrm{T}}}}
\newcommand{\nRx}{{n_{\mathrm{R}}}}
\newcommand{\TxCM}{\B{\Psi}_{\mathrm{T}}}
\newcommand{\RxCM}{\B{\Psi}_{\mathrm{R}}}
\newcommand{\TxCC}{{\rho_{\mathrm{T}}}}
\newcommand{\RxCC}{{\rho_{\mathrm{R}}}}

\newcommand{\nXmin}{{n_\mathrm{S}}}
\newcommand{\nXmax}{{n_\mathrm{L}}}

\newcommand{\CMmin}{\B{\Psi}_{\mathrm{S}}}
\newcommand{\CMmax}{\B{\Psi}_{\mathrm{L}}}
\newcommand{\Emin}[1]{\lambda_{\mathrm{S},{#1}}}
\newcommand{\Emax}[1]{\lambda_{\mathrm{L},{#1}}}
\newcommand{\Kc}{K_{\mathrm{cor}}}

\newcommand{\MC}{\B{\Lambda}}
\newcommand{\MI}{\B{\Omega}}
\newcommand{\MHH}{\B{\Theta}}
\newcommand{\snr}{\eta}
\newcommand{\ExpCM}[2]{\B{\Phi}_{#1}^{\left(\mathrm{exp}\right)}\left(#2\right)}

\newcommand{\mySep}{\vspace*{5pt}}

\begin{document}
\title{
        \hspace{2cm}\\[1cm]
        On the Capacity of Doubly Correlated MIMO Channels
      }
\author{
        \hspace{4cm}\\[-0.5cm]
        {\normalsize
        Hyundong Shin, \IEEEmembership{\normalsize Member, IEEE},
        Moe Z. Win, \IEEEmembership{\normalsize Fellow, IEEE},\\
        Jae Hong Lee, \IEEEmembership{\normalsize Senior Member, IEEE},
        and
        Marco Chiani, \IEEEmembership{\normalsize Senior Member, IEEE} \vspace*{7mm}\\
        }
\normalsize{
\underline{Corresponding Address:} \\[3pt]
Hyundong Shin\\
Laboratory for Information and Decision Systems (LIDS) \\
Massachusetts Institute of Technology \\
Cambridge, MA 02139, USA \\
Tel.: (617) 253-6173 \\
e-mail: {\tt hshin@mit.edu} \\
\thanks{H. Shin and M. Win are with
            the Laboratory for Information and Decision Systems (LIDS),
            Massachusetts Institute of Technology,
            Room 32-D658, 77 Massachusetts Avenue,
            Cambridge, MA 02139, USA
            (e-mail: {\tt hshin@mit.edu},
                {\tt moewin@mit.edu}).}
\thanks{J. H. Lee is with
            the School of Electrical Engineering,
            Seoul National University, Seoul, Korea
            (e-mail: {\tt jhlee@snu.ac.kr}).} %
\thanks{M. Chiani is with IEIIT-BO/CNR, DEIS, University of Bologna,
            Viale Risorgimento 2,
            40136 Bologna, Italy
            (e-mail: {\tt mchiani@deis.unibo.it}).
       }
} }

\markboth{IEEE Transactions on Wireless Communications, vol. XX,
no. X, month 2005}
         {Shin \textit{\MakeLowercase{et al.}}:
          On the Capacity of Doubly Correlated MIMO Channels}

\maketitle

\clearpage

\begin{abstract}
In this paper, we analyze the capacity of multiple-input
multiple-output (MIMO) Rayleigh-fading channels in the presence of
spatial fading correlation at \emph{both} the transmitter and the
receiver, assuming that the channel is unknown at the transmitter
and perfectly known at the receiver. We first derive the
\emph{determinant} representation for the exact characteristic
function of the capacity, which is then used to determine the
\emph{trace} representations for the mean, variance, skewness,
kurtosis, and other higher-order statistics (HOS). These results
allow us to exactly evaluate two relevant information-theoretic
capacity measures---ergodic capacity and outage capacity---and the
HOS of the capacity for such a MIMO channel. The analytical
framework presented in the paper is valid for arbitrary numbers of
antennas and generalizes the previously known results for
independent and identically distributed or one-sided correlated
MIMO channels to the case when fading correlation exists on both
sides. We verify our analytical results by comparing them with
Monte Carlo simulations for a correlation model based on realistic
channel measurements as well as a classical exponential
correlation model.
\end{abstract}


\section{Introduction}

Multiple-input multiple-output (MIMO) communication systems using
multiple transmit and receive antennas promise high spectral
efficiency and link reliability for wireless communications
\cite{Wi87,FG98,Te99}. Although the linear growth of capacity with
the number of antennas indicates the potential of MIMO systems,
the true benefits of the use of multiple antennas may be limited
by spatial fading correlation due to closely-spaced antenna
configurations and poor scattering environments in realistic
wireless channels \cite{SFGK00,CTKV02}.

Since the pioneering work of \cite{Wi87,FG98,Te99} in the area of
multiple-antenna communications predicted remarkable spectral
efficiency of MIMO wireless systems in independent and identically
distributed (i.i.d.) Rayleigh fading, much subsequent work has
concentrated on characterizing MIMO capacity under correlated
fading
\cite{SFGK00,CTKV02,SL03IT,CWZ03IT,SRS03,KA03,SGL03,GSSC03,ONBP03,MSS03,MO04}.
However, the exact analytical results for the capacity such as
ergodic (or mean) capacity, capacity variance, and outage capacity
(i.e., capacity versus outage probability)\footnote{
        In general, the capacity distribution is required to determine
        the outage capacity \cite{FG98,Te99,BPS98}.
        }
have been known for only a few special cases, largely due to
mathematical intractability (see, e.g., \cite{Te99,SL03IT,SS04}
for i.i.d.\ flat Rayleigh fading and \cite{CWZ03IT,SRS03,KA03} for
a one-sided correlated MIMO channel). For a more general case of
correlated fading at both the transmitter and the receiver, which
we will refer to as \emph{doubly correlated} MIMO channels in the
paper, some limited results are available: the capacity
distribution for a small number of antennas (i.e., $\min
\left\{\nTx,\nRx\right\} \leq 3$ where $\nTx$ and $\nRx$ are the
numbers of transmit and receive antennas, respectively)
\cite{SGL03}, upper and lower bounds on the ergodic capacity
\cite{SL03IT,ONBP03}, capacity statistics for the case with a
large number of antennas \cite{MSS03}, and the asymptotic mean and
variance of the capacity in the limit as the number of antennas
tends to infinity \cite{CTKV02,MO04}. The temporal behavior of the
capacity was analyzed in \cite{GSSC03} in terms of level crossing
rates and average fade durations.

In this paper, we focus on deriving the exact analytical
expressions for capacity statistics of doubly correlated MIMO
Rayleigh-fading channels using the methodology developed in
\cite{SL03IT} and \cite{CWZ03IT}, assuming perfect channel
knowledge at the receiver and no knowledge at the transmitter with
the average input-power constraint. The principal contributions of
this paper are as follows.

\begin{itemize}

\item We derive a \emph{determinant} representation for the
characteristic function (CF) of MIMO capacity, which generalizes
the previous results for i.i.d.\ and one-sided correlated channels
\cite{CWZ03IT,SRS03,KA03} to the doubly correlated case.

\item We derive \emph{trace} representations for the mean,
variance, and higher-order statistics (HOS) (e.g., cumulants,
skewness, and kurtosis) of the capacity using the determinant
representation of the CF and the relationship between
\emph{polymatrices} and \emph{dimatrices}.\footnote{
        The definitions of the polymatrix and the dimatrix will be
        introduced in Section \ref{sec:Sec3}.
        }

\item We characterize the effect of fading correlation on the
capacity statistics at high signal-to-noise ratio (SNR). We show
that at high SNR, the variance, skewness, kurtosis, and other HOS
of the capacity depend only on correlation at the side with the
larger number of antennas. Moreover, when $\nTx=\nRx$, these
statistics are not affected by fading correlation at any side.

\end{itemize}
To verify our analytical results, we also compare them with Monte
Carlo simulations for doubly correlated MIMO channels using a
correlation model based on physical measurements
\cite{Ke02,C802.20-03/50} as well as a classical exponential
correlation model. It should be noted that alternative derivation
of the moment generating function (MGF) of the capacity for doubly
correlated MIMO channels can also be found in \cite{Ki05}. In this
study, the MGF was obtained \emph{indirectly} from the case of a
square channel matrix ($\nTx=\nRx$) using the \emph{limiting}
approach of \cite{GS00} and then, the first moment (ergodic
capacity) was deduced from it in terms of a sum of
$\min\left\{\nTx,\nRx\right\}$ determinants.

The remainder of the paper is organized as follows. A brief
overview of the distributions of complex random matrices required
for our analysis, channel model, and associated channel capacity
are presented in Section \ref{sec:Sec2}. The CF of the capacity is
derived and the capacity statistics are analyzed for doubly
correlated MIMO channels in Section \ref{sec:Sec3}. The effect of
fading correlation on the capacity statistics is investigated at
high SNR in Section \ref{sec:Sec4}. In Section \ref{sec:Sec5},
some numerical and simulation results are provided to illustrate
our analytical results. Finally, Section \ref{sec:Sec6} concludes
the paper.

We shall use the following notation throughout the paper.
$\mathbb{N}$ and $\C$ denote the natural numbers and the field of
complex numbers, respectively. The superscript $\dag$ denotes the
transpose conjugate. $\B{I}_n$ and $\tr\left(\B{A}\right)$
represent the $n\times n$ identity matrix and the trace operator
of a square matrix $\B{A}$, respectively. By $\B{A}>0$, we denote
that $\B{A}$ is positive definite. For a matrix
$\B{A}\left(t\right)=\left[a_{i,j}\left(t\right)\right]$ where
$a_{i,j}\left(t\right)$ are differentiable functions of $t$, the
$n$th derivative of $\B{A}\left(t\right)$ with respect to $t$ is
denoted by $\B{A}^{\left(n\right)}\left(t\right)=\left[d^n
a_{i,j}\left(t\right)/dt^n\right]$.

\section{Preliminaries: Definitions and Models}     \label{sec:Sec2}

In this section, we give a brief overview of the distribution
theory of complex random matrices (that serves as a central
mathematical tool to analyze MIMO communication systems), channel
model, and associated channel capacity.

\subsection{Distributions of Complex Random Matrices}

Let us denote a complex Gaussian matrix $\B{X} \in \C^{m \times
n}$ with the probability density function (PDF)
\cite[eq.~(1)]{SL03IT}
\begin{align}    \label{eq:CGM}
    \PDF{\B{X}}{\B{X}}
        =
            \pi^{-mn}
            \det\left(\B{\Sigma}\right)^{-n}
            \det\left(\B{\Psi}\right)^{-m}
            e^{
                -\tr\left\{
                    \B{\Sigma}^{-1}
                    \left(\B{X}-\B{M}\right)
                    \B{\Psi}^{-1}
                    \left(\B{X}-\B{M}\right)^\dag
                \right\}
            }
\end{align}
by $\B{X} \sim \CGM{m}{n}{\B{M}}{\B{\Sigma}}{\B{\Psi}}$ where
$\B{\Sigma} \in \C^{m \times m}>0$ and $\B{\Psi} \in \C^{n \times
n}>0$ are Hermitian. If $\B{X} \sim
\CGM{m}{n}{\B{0}}{\B{\Sigma}}{\B{I}_n}$, $m\leq n$, and
$\B{Y}=\B{XX}^\dag$, then $\B{Y}$ has a complex (central) Wishart
density $\CWM{m}{n}{\B{\Sigma}}$ given by \cite[eq.~(3)]{SL03IT}.

\mySep

\begin{definition}[Quadratic Form in the Complex Gaussian Matrix \cite{SL03IT}] \label{def:CQM}
Let
$$\B{X} \sim \CGM{m}{n}{\B{0}}{\B{\Sigma}}{\B{\Psi}},
\quad m\leq n.$$
A positive-definite quadratic form $\B{Y}$ in $\B{X}$ associated
with a Hermitian matrix $\B{A} \in \C^{n \times n}>0$, denoted by
$\B{Y} \sim \CQM{m}{n}{\B{A}}{\B{\Sigma}}{\B{\Psi}}$, is then
defined as $\B{Y}=\B{XAX}^\dag$.
\end{definition}

\mySep

The PDF of $\B{Y} \sim \CQM{m}{n}{\B{A}}{\B{\Sigma}}{\B{\Psi}}$ is
given by \cite[eq.~(57)]{Kh66} and can be expressed in an
equivalent form
\begin{align}    \label{eq:CQM}
    \PDF{\B{Y}}{\B{Y}}
    &=      \frac{1}{\tilde{\Gamma}_m\left(n\right)}
            \det\left(\B{\Sigma}\right)^{-n}
            \det\left(\B{A}\B{\Psi}\right)^{-m}
            \det\left(\B{Y}\right)^{n-m}
            {}_0\tilde{F}_0^{\left(n\right)}\left(
                -\B{\Sigma}^{-1}\B{Y},
                \B{\Psi}^{-1}\B{A}^{-1}
            \right),
        ~
        \B{Y}>0
\end{align}
where $\tilde {\Gamma }_m \left( \alpha \right)=\pi^{m\left( m-1
\right)/ 2}\prod\nolimits_{i=0}^{m-1} {\Gamma \left( {\alpha -i}
\right)}$, $\mathrm{Re}\left( \alpha \right)>m-1$, is the complex
multivariate gamma function, $\Gamma \left( \cdot \right)$ is the
gamma function, and
${}_p\tilde{F}_q^{\left(n\right)}\left(\cdot\right)$ is the
hypergeometric function of two Hermitian matrices, defined by
\cite[eq.~(51)]{Kh66}. Note that the density \eqref{eq:CQM} is a
counterpart of the real case in \cite[eq.~(7.2.5)]{GN00} and if
$\B{A}\B{\Psi}=\B{I}_n$, it reduces to the complex Wishart density
$\CWM{m}{n}{\B{\Sigma}}$.

\subsection{Channel Model and Capacity Random Variable}

We consider a point-to-point frequency-flat fading MIMO link with
$\nTx$ transmit and $\nRx$ receive antennas. Let $\B{x} \in
\C^{\nTx}$ be a transmitted signal vector with input covariance
$\B{Q}=\EX{\B{xx}^\dag}$ satisfying the power constraint $\tr
\left(\B{Q}\right) \leq \mathcal{P}$, then the received signal is
given by
\begin{align}    \label{eq:RS}
    \B{y}=\B{H}\B{x} + \B{n}
\end{align}
where $\B{H}\in \C^{\nRx \times \nTx}$ is the random channel
matrix whose $\left(i,j\right)$th entries $H_{ij}$, $i=1,2,\ldots
,\nRx$, $j=1,2,\ldots ,\nTx$, are complex propagation coefficients
between the $j$th transmit antenna and the $i$th receive antenna
with $\EX{|H_{ij}|^2}=1$, and $\B{n}$ is the complex
$\nRx$-dimensional zero-mean additive white Gaussian noise (AWGN)
vector with the covariance matrix $\sigma_n^2 \B{I}_{\nRx}$.
For doubly correlated
MIMO channels, the channel matrix $\B{H}$ can be written as
\cite{SFGK00,CTKV02}
\begin{align}    \label{eq:H}
    \B{H}=\RxCM^{1/2} \B{H}_\text{uc} \TxCM^{1/2}
\end{align}
where $\B{H}_\text{uc} \sim
\CGM{\nRx}{\nTx}{\B{0}}{\B{I}_\nRx}{\B{I}_\nTx}$, and $\TxCM \in
\C^{\nTx \times \nTx}>0$ and $\RxCM \in \C^{\nRx \times \nRx}>0$
are the transmit and receive correlation matrices, respectively.
Note that $\B{H} \sim \CGM{\nRx}{\nTx}{\B{0}}{\RxCM}{\TxCM}$ and
it has been used in various attempts for studying correlated MIMO
channels
\cite{SFGK00,CTKV02,SL03IT,CWZ03IT,SRS03,KA03,SGL03,GSSC03,ONBP03,MSS03,MO04}.
Recently, this model has also been validated through physical
measurements \cite{Ke02}.

In what follows, we refer to $\nXmin=\min\left\{\nTx,\nRx\right\}$
and $\nXmax=\max\left\{\nTx,\nRx\right\}$ and define the
random matrix $\MHH \in \C^{\nXmin \times \nXmin}>0$ as
\begin{align*}
    \MHH \triangleq
        \begin{cases}
            \B{HH}^\dag,        &   \text{if } \nRx \leq \nTx
            \\[-2mm]
            \B{H}^\dag \B{H},   &   \text{otherwise}.
        \end{cases}
\mySep
\end{align*}
Also, let us denote, for convenience,
\begin{align*}
    \left(\CMmin,\CMmax\right) =
        \begin{cases}
            \left(\RxCM,\TxCM\right),   &   \text{if } \nRx \leq \nTx
            \\[-2mm]
            \left(\TxCM,\RxCM\right),   &   \text{otherwise}
        \end{cases}
\end{align*}
and let $0<\Emin{1}<\Emin{2}<\cdots<\Emin{\nXmin}$ and
$0<\Emax{1}<\Emax{2}<\cdots<\Emax{\nXmax}$ be distinct ordered
eigenvalues of $\CMmin$ and $\CMmax$, respectively. Then, $\MHH
\sim \CQM{\nXmin}{\nXmax}{\B{I}_\nXmax}{\CMmin}{\CMmax}$ for the
doubly correlated MIMO channel.


In general, when the receiver has perfect channel knowledge, the
input distribution that maximizes the mutual information between
$\B{x}$ and $\B{y}$ is circularly symmetric complex Gaussian for
any given input covariance $\B{Q}$. When the transmitter has no
channel knowledge, power among transmit antennas cannot be
allocated in accordance with the realization of $\B{H}$ to
maximize the mutual information, and hence equal power allocation
to each of transmit antennas is the most reasonable strategy,
i.e., choosing $\B{Q}=\left(\mathcal{P}/\nTx\right)\B{I}_\nTx
$.\footnote{
        It has been shown in \cite{Te99} that if the channel
        has i.i.d.\ Rayleigh fading between antenna pairs, the optimum
        input covariance matrix is $\B{Q}=\left(\mathcal{P}/\nTx\right)
        \B{I}_\nTx$.
        }
This yields the capacity in nats/s/Hz as \cite{FG98,Te99}
\begin{align}    \label{eq:C}
    C = \ln \det\left(
                \B{I}_\nXmin +
                \left(\snr/\nTx\right)\MHH
                \right)
\end{align}
where $\snr=\mathcal{P}/\sigma_n^2$ is the average SNR at each
receive antenna. Since the channel matrix $\B{H}$ is random, the
associated channel capacity $C$ is also a random variable whose
statistics are determined by the statistical properties of the
eigenvalues of $\MHH \sim
\CQM{\nXmin}{\nXmax}{\B{I}_\nXmax}{\CMmin}{\CMmax}$.

\section{Capacity Statistics}   \label{sec:Sec3}

In this section, we will investigate the statistical properties of
the capacity random variable $C$ in \eqref{eq:C} for doubly
correlated MIMO channels. We begin by deriving the CF\ of $C$,
from which all other functions, such as the PDF, cumulative
distribution function (CDF), and cumulant generating function
(CGF), and statistical moments of $C$ can be obtained.

\subsection{Characteristic Function}    \label{sec:Sec3-A}

\begin{theorem}     \label{th:CFCcor}

Let $\B{H} \sim \CGM{\nRx}{\nTx}{\B{0}}{\RxCM}{\TxCM}$, i.e.,
$\MHH \sim \CQM{\nXmin}{\nXmax}{\B{I}_\nXmax}{\CMmin}{\CMmax}$.
Then, the CF of the capacity $C$ in nats/s/Hz is
\begin{align}    \label{eq:CFCcor}
    \CF{C}{\jmath\omega}
    & \triangleq
        \EX{e^{\jmath\omega C}}
    \nonumber \\
    &=
        \Kc^{-1} \,
        \Upsilon_\nXmin\left(\jmath\omega\right) \,
        \det
                \MC\left(\jmath\omega\right)
\end{align}
where $\jmath=\sqrt{-1}$ and
\begin{align}    \label{eq:Kc}
    \Kc =   \left(\frac{\snr}{\nTx}\right)^{\nXmin\left(\nXmin-1\right)/2}
            \prod_{1\leq i<j\leq \nXmin} \left(\Emin{j}-\Emin{i}\right)
            \prod_{1\leq i<j\leq \nXmax} \left(\Emax{j}-\Emax{i}\right)
\end{align}
\begin{align}    \label{eq:Um}
    \Upsilon_\nXmin \left(\jmath\omega\right)
        =   \prod_{\ell=1}^{\nXmin-1}
                \left(\jmath\omega+\ell\right)^{-\ell}
\end{align}
and $\MC\left(\jmath\omega\right)$ is the $\nXmax \times \nXmax$
matrix whose $\left(i,j\right)$th entry is given in Table
\ref{table:Table1}.

\end{theorem}

\begin{proof}
See Appendix B.
\end{proof}

\mySep

Note that Theorem \ref{th:CFCcor} requires correlation matrices
$\TxCM$ and $\RxCM$ to have distinct eigenvalues. The case when
the correlation matrices have non-distinct eigenvalues (some of
$\Emin{i}$'s or $\Emax{i}$'s are equal), we can obtain the CF as a
limiting case of \eqref{eq:CFCcor} \cite{CWZ03IT,Kh70}. In
particular, when $\TxCM=\B{I}_{\nTx}$ and $\RxCM=\B{I}_{\nRx}$
(i.i.d.\ case), $\CF{C}{\jmath\omega}$ is given by
\cite[eq.~(25)]{CWZ03IT}
\begin{align}    \label{eq:CFCiid}
    \CF{C}{\jmath\omega}
        =   K_{\mathrm{iid}}^{-1} \,
            \det
                    \MI\left(\jmath\omega\right)
\end{align}
where $K_{\mathrm{iid}} =\prod_{\ell=1}^\nXmin
\left(\nXmax-\ell\right)! \left(\ell-1\right)!$ and
$\MI\left(\jmath\omega\right)$ is the $\nXmin \times \nXmin$
Hankel matrix whose $\left(i,j\right)$th entry is given in Table
\ref{table:Table1}. 
%
%
Theorem \ref{th:CFCcor} generalizes the previous results for
i.i.d.\ and one-sided correlated channels (which are special cases
of non-distinct eigenvalues) \cite{CWZ03IT,SRS03,KA03} to the
doubly correlated MIMO channel given by \eqref{eq:H}. Using the
analytical formulas for the CF\ in \eqref{eq:CFCcor} and
\eqref{eq:CFCiid}, the PDF and CDF of $C$ can be expressed in
forms of the inverse Fourier transform of $\CF{C}{\jmath\omega}$,
%
%
%
which can be efficiently calculated by using the fast Fourier
transform (FFT) method \cite{CWZ03IT,Sh04}.

\subsection{Mean, Variance, and Higher-Order Statistics}

From the CF\ of $C$ in \eqref{eq:CFCcor} and \eqref{eq:CFCiid}
involving the determinants, we derive the exact closed-form
expressions for the mean, variance, and other HOS such as
cumulants, skewness, and (excess) kurtosis of the capacity. To do
this, we first introduce the logarithmic derivative of a
determinant.

\subsubsection{Logarithmic Derivatives of a Determinant}

Let $\B{R}\left(t\right)$ be a matrix depending on a parameter
$t$. If each entry of $\B{R}\left(t\right)$ is differentiable with
respect to $t$, then so is $\det \B{R}\left(t\right)$ because the
determinant is a polynomial in the entries of
$\B{R}\left(t\right)$. If $\B{R}\left(t\right)$ is invertible, the
first-order logarithmic derivative of $\det \B{R}\left(t\right)$
is given by \cite{Go72}
\begin{align}    \label{eq:LDD1}
    \frac{d \ln\det\B{R}\left(t\right)}{dt}
        = \tr\left\{
                \B{R}^{-1}\left(t\right)
                \B{R}^{\left(1\right)}\left(t\right)
            \right\}.
\end{align}
We now generalize \eqref{eq:LDD1} to the arbitrary order of
differentiation.

\mySep

\begin{definition}[Polymatrix and Dimatrix]     \label{def:PM}

Let $\B{R}\left(t\right)$ be an invertible matrix whose elements
are differentiable with respect to $t$. Then, the $n$th
\emph{polymatrix} of $\B{R}\left(t\right)$ with respect to $t$ is
defined as
\begin{align}    \label{eq:PM}
    \B{R}_{\left[n\right]} \left(t\right) \triangleq
        \B{R}^{-1}\left(t\right)
        \B{R}^{\left(n\right)}\left(t\right).
\end{align}
In particular, we call $\B{R}_{\left[1\right]} \left(t\right)$ the
\emph{dimatrix} of $\B{R}\left(t\right)$.
\end{definition}

\mySep

\begin{lemma}   \label{le:LDD}

The polymatrices and the derivatives of the dimatrix of
$\B{R}\left(t\right)$ have the following relationship
\begin{align}    \label{eq:RPD}
    \B{R}_{\left[n\right]} \left(t\right)
        =   \sum_{\ell=1}^n
                \binom{n-1}{\ell-1}
                \B{R}_{\left[n-\ell\right]} \left(t\right)
                \B{R}_{\left[1\right]}^{\left(\ell-1\right)} \left(t\right)
\end{align}
and the $\ell$th logarithmic derivative of $\det
\B{R}\left(t\right)$ is the trace of the $\left(\ell-1\right)$th
derivative of the dimatrix $\B{R}_{\left[1\right]}
\left(t\right)$, i.e.,
\begin{align}    \label{eq:LDD}
    \frac{d^\ell \ln\det \B{R}\left(t\right)}{dt^\ell}
        =   \tr\left\{
                    \B{R}_{\left[1\right]}^{\left(\ell-1\right)}
                    \left(t\right)
            \right\}.
\end{align}
\end{lemma}

\mySep

\begin{proof}
By definition, we have
\begin{align*}
    \B{R}_{\left[n\right]} \left(t\right)
        &   =   \B{R}^{-1}\left(t\right)
                \frac{d^{n-1} \B{R}^{\left(1\right)}\left(t\right)}{dt^{n-1}}
        \\
        &   \stackrel{\left(a\right)}{=}
                \B{R}^{-1}\left(t\right)
                \sum_{\ell=1}^n
                    \binom{n-1}{\ell-1}
                    \frac{d^{n-\ell}\B{R}\left(t\right)}{dt^{n-\ell}}
                    \frac{d^{\ell-1}\B{R}_{\left[1\right]}\left(t\right)}
                        {dt^{\ell-1}}
        \\
        &   \stackrel{\left(b\right)}{=}
                \sum_{\ell=1}^n
                    \binom{n-1}{\ell-1}
                    \B{R}_{\left[n-\ell\right]}\left(t\right)
                    \B{R}_{\left[1\right]}^{\left(\ell-1\right)}
                    \left(t\right)
\end{align*}
where (a) follows from the Leibniz's identity \cite[p. 21]{GR00}
and (b) follows from \eqref{eq:PM}. Also, \eqref{eq:LDD} follows
immediately from \eqref{eq:LDD1}, \eqref{eq:PM}, and by
interchanging the order of differentiation and trace operators.
\end{proof}

\mySep

This lemma says that the $\ell$th logarithmic derivative of the
determinant of a matrix can be determined by its first $\ell$
polymatrices. For example, the second, third, and fourth order
logarithmic derivatives of $\det \B{R}\left(t\right)$ are given by
\begin{align}
    \label{eq:LDD2}
    \frac{d^2 \ln\det \B{R}\left(t\right)}{dt^2}
    &   =   \tr\left\{
                    \B{R}_{\left[2\right]}\left(t\right)
                    -\B{R}_{\left[1\right]}^2 \left(t\right)
                \right\}
    \\
    \label{eq:LDD3}
    \frac{d^3 \ln\det \B{R}\left(t\right)}{dt^3}
    &   =   \tr\left\{
                    2\B{R}_{\left[1\right]}^3 \left(t\right)
                    -3\B{R}_{\left[1\right]} \left(t\right)
                    \B{R}_{\left[2\right]} \left(t\right)
                    +\B{R}_{\left[3\right]} \left(t\right)
                \right\}
    \\
    \label{eq:LDD4}
    \frac{d^4 \ln\det \B{R}\left(t\right)}{dt^4}
    &   =   \tr\left\{
                    -6\B{R}_{\left[1\right]}^4 \left(t\right)
                    +12\B{R}_{\left[1\right]}^2 \left(t\right)
                    \B{R}_{\left[2\right]} \left(t\right)
                \right.
    \nonumber \\
    & \qquad \quad
                \left.
                    -3\B{R}_{\left[2\right]}^2 \left(t\right)
                    -4\B{R}_{\left[1\right]} \left(t\right)
                    \B{R}_{\left[3\right]} \left(t\right)
                    +\B{R}_{\left[4\right]} \left(t\right)
                \right\}.
\end{align}
Using the explicit determinantal CF's in Section \ref{sec:Sec3-A}
and Lemma \ref{le:LDD}, we now derive statistical moments of $C$,
which requires determining the polymatrices
\begin{align}    \label{eq:PMiid}
    \MI_{\left[n\right]}\left(\nu\right) =
        \MI^{-1}\left(\nu\right)
        \MI^{\left(n\right)}\left(\nu\right),
    \quad
    \MI_{\left[0\right]}\left(\nu\right)=\B{I}_\nXmin
\end{align}
%
%
%
%
\begin{align}    \label{eq:PMcor}
    \MC_{\left[n\right]}\left(\nu\right) =
        \MC^{-1}\left(\nu\right)
        \MC^{\left(n\right)}\left(\nu\right),
    \quad
    \MC_{\left[0\right]}\left(\nu\right)=\B{I}_\nXmax
\end{align}
for i.i.d.\ and doubly correlated MIMO channels, respectively,
where the $(i,j)$th entries of
$\MI^{\left(n\right)}\left(\nu\right)$ and
$\MC^{\left(n\right)}\left(\nu\right)$ for $n \in \mathbb{N}$ are
given in Table \ref{table:Table1}.

\mySep

\subsubsection{Cumulants}

The $n$th cumulant of $C$ is by definition expressed as
\begin{align}    \label{eq:CumC}%
    \kappa_{n} \triangleq
        \left.
            \frac{d^n}{d\nu^n}
            \CGF{C}{\nu}
        \right|_{\nu=0}
\end{align}
where $\CGF{C}{\nu} \triangleq \ln\CF{C}{\nu}$ is the CGF of $C$.
Note that the first and second cumulants are the mean and variance
of the capacity, respectively.

\mySep

\begin{theorem}     \label{th:CumCiid}

Let $\B{H} \sim \CGM{\nRx}{\nTx}{\B{0}}{\B{I}_\nRx}{\B{I}_\nTx}$,
i.e., $\MHH \sim \CWM{\nXmin}{\nXmax}{\B{I}_\nXmin}$ ($\nRx \times
\nTx$ i.i.d.\ MIMO channel). Then, the $n$th cumulant of the
capacity $C$ in nats/s/Hz is
\begin{align}    \label{eq:CumCiid}%
    \kappa_{n}
    =   \tr\left\{
                \MI_{\left[1\right]}^{\left(n-1\right)} \left(0\right)
            \right\}.
\mySep
\end{align}

\end{theorem}

\mySep

\begin{proof}
It follows immediately from \eqref{eq:CFCiid}, \eqref{eq:CumC},
and Lemma \ref{le:LDD}.
\end{proof}

\mySep

\begin{theorem}     \label{th:CumCcor}

Let $\B{H} \sim \CGM{\nRx}{\nTx}{\B{0}}{\RxCM}{\TxCM}$, i.e.,
$\MHH \sim \CQM{\nXmin}{\nXmax}{\B{I}_\nXmax}{\CMmin}{\CMmax}$
($\nRx \times \nTx$ doubly correlated MIMO channel). Then, the
$n$th cumulant of the capacity $C$ in nats/s/Hz is
\begin{align}    \label{eq:CumCcor}%
    \kappa_{n}
    =   \tr\left\{
                \MC_{\left[1\right]}^{\left(n-1\right)}
                \left(0\right)
            \right\}
        +\left(-1\right)^n
        \left(n-1\right)!
        \sum_{\ell=1}^{\nXmin-1}
            \ell^{-n+1}.
\mySep
\end{align}

\end{theorem}

\mySep

\begin{proof}
It follows immediately from Theorem \ref{th:CFCcor},
\eqref{eq:CumC}, and Lemma \ref{le:LDD}.
\end{proof}

\mySep

Using the relationship \eqref{eq:RPD}, the matrices
$\MI_{\left[1\right]}^{\left(n-1\right)} \left(\nu\right)$ and
$\MC_{\left[1\right]}^{\left(n-1\right)} \left(\nu\right)$ in
Theorems \ref{th:CumCiid} and \ref{th:CumCcor} can be determined
by the polymatrices in \eqref{eq:PMiid} and \eqref{eq:PMcor},
respectively.

\mySep

\subsubsection{Raw and Central Moments}

The $n$th raw and central moments (i.e., moments about the origin
and the mean, respectively) of $C$ can be obtained from the
general relationships between moments and cumulants \cite{EHP00}:
\begin{align}
    \label{eq:CRM}
    m_{n}
       \triangleq
        \EX{C^n} =
            \sum_{\ell=1}^n
                \binom{n-1}{\ell-1}
                m_{n-\ell} \,
                \kappa_{\ell}
\end{align}
\begin{align}
    \label{eq:CCM}
    \mu_{n}
       \triangleq
        \EX{\left(C-m_{1}\right)^n} =
            \sum_{\ell=0}^n
                \binom{n}{\ell}
                m_{n-\ell}
                \left(-m_{1}\right)^\ell
\end{align}
where $m_{n}$ and $\mu_{n}$ are the $n$th raw and central moments
of $C$, respectively. In particular, the ergodic capacity is given
by $m_1=\EX{C}$.

\subsubsection{Skewness and Kurtosis}

The skewness characterizes the degree of asymmetry of a
distribution around its mean and the kurtosis measures the
peakedness or flatness of a distribution relative to a Gaussian
distribution. The skewness and (excess) kurtosis of $C$, denoted
by $\beta_{1}$ and $\beta_{2}$, respectively, can be obtained from
the cumulants in Theorems \ref{th:CumCiid} and \ref{th:CumCcor}
for i.i.d.\ and doubly correlated cases as
\begin{align}    \label{eq:SkewC}
    \beta_{1} \triangleq
        \frac{\mu_{3}}{\mu_{2}^{3/2}} =
        \frac{\kappa_{3}}{\kappa_{2}^{3/2}}
\end{align}
\begin{align}    \label{eq:KurtC}
    \beta_{2} \triangleq
        \frac{\mu_{4}}{\mu_{2}^2}-3 =
        \frac{\kappa_{4}}{\kappa_{2}^2}.
\end{align}

Using \eqref{eq:LDD2}--\eqref{eq:LDD4}, \eqref{eq:CumCiid},
\eqref{eq:CumCcor}, \eqref{eq:SkewC}, and \eqref{eq:KurtC}, the
trace representations for the mean, variance, skewness, and
kurtosis of the capacity are tabulated in Tables
\ref{table:Table2} and \ref{table:Table3} for i.i.d.\ and doubly
correlated MIMO channels, respectively.\footnote{
        Alternative expressions for the ergodic capacity and capacity
        variance of the i.i.d.\ case, in terms of integrals involving the Laguerre
        polynomials, can be found in
        \cite{Te99} and \cite{SS04}, respectively.
        }
Since the skewness and kurtosis of a Gaussian distribution are
equal to zero, nonzero values of these quantities indicate the
degree of deviation from the Gaussian distribution.\footnote{
        The skewness and kurtosis of the capacity have been used
        in conjunction with the Gram-Charlier expansion
        to estimate the Kullback-Leiblier divergence, as a measure
        of non-Gaussianity, between the capacity distribution and
        its Gaussian approximation \cite{Sh04}.
        }

\section{Effect of Correlation at High SNR}     \label{sec:Sec4}

The effect of fading correlation on the behavior of capacity is
not immediately apparent from the exact analytical expressions
such as the CF\ in \eqref{eq:CFCcor} and the moments in Table
\ref{table:Table3}. Therefore, we resort to the asymptotic
analysis to investigate such an effect in the following. In
particular, we consider a high-SNR regime since the benefits of
the use of multiple antennas are more pronounced at high SNR. In
this case, the capacity \eqref{eq:C} can be written as
\begin{align}    \label{eq:ChSNR}%
    C =
        \nXmin \ln\left(\snr/\nTx\right)
        +\ln\det\left(\MHH\right)
        +O\left(1/\snr\right),
\end{align}
which reveals that at high SNR, the capacity is characterized by
the logarithmic generalized variance $\ln\det\left(\MHH\right)$ of
$\MHH \sim \CQM{\nXmin}{\nXmax}{\B{I}_\nXmax}{\CMmin}{\CMmax}$.
Note in \eqref{eq:ChSNR} that $\min\left\{\nTx,\nRx\right\}$ (more
precisely, the rank of $\B{H})$ determines the spatial
multiplexing gain of a MIMO channel, while the logarithmic
generalized variance $\ln\det\left(\MHH\right)$ determines the
diversity gain in capacity point of view. Consequently, if the
correlation matrices have full rank, then the channel matrix
$\B{H}$ in \eqref{eq:H} has full rank with probability one. In
this case, antenna correlation does not diminish the spatial
multiplexing gain and only decreases the diversity gain
\cite{SL03IT}.

Starting with \eqref{eq:ChSNR} and using similar steps in the
proof of Theorem \ref{th:CFCcor}, the CF\ of the capacity in the
high-SNR regime can be written as
\begin{align}   \label{eq:CFChSNR}
    \CF{C}{\jmath\omega}
        \big|_{\snr ~\text{high}}
        & =
            A
            \int_{0<z_1 \leq \cdots \leq z_\nXmin <\infty}
                \prod_{\ell=1}^\nXmin
                    z_\ell^{\jmath\omega}
                \prod_{1\leq i<j\leq \nXmin}
                    \left(z_j -z_i\right) \;
                \det\left(\B{\Xi}\right)
                dz_1 dz_2 \cdots dz_\nXmin
        \nonumber \\
        & =
            A
            \left\{
                \prod_{\ell=1}^\nXmin
                    \Gamma\left(\jmath\omega+\ell\right)
            \right\} \,
            \det
                    \B{K}\left(\jmath\omega\right)
\end{align}
where
\begin{align}    \label{eq:A}
    A =
        \frac{
                \left(\snr/\nTx\right)^{\jmath\omega\nXmin}
                \det\left(\CMmin\right)^{\jmath\omega}
            }
            {
                \prod_{\ell=1}^\nXmin \Gamma\left(\ell\right)
                \prod_{1\leq i<j\leq \nXmax}
                    \left(\Emax{j}-\Emax{i}\right)
            }
\end{align}
and $\B{K}\left(\jmath\omega\right)$ is the $\nXmax \times \nXmax$
matrix whose $(i,j)$th entry is given by
\begin{align}    \label{eq:MK}
    \left\{\B{K}\left(\jmath\omega\right)\right\}_{i,j} =
        \begin{cases}
            \Emax{j}^{i-1}, &
                i=1,\ldots,\nXmax-\nXmin, ~j=1,\ldots,\nXmax
            \\
            \Emax{j}^{\jmath\omega+i-1}, &
                i=\nXmax-\nXmin+1,\ldots,\nXmax,
                ~j=1,\ldots,\nXmax.
        \end{cases}
\end{align}
Also, from Lemma \ref{le:LDD}, \eqref{eq:CumC}, and
\eqref{eq:CFChSNR}, the $n$th cumulant of the capacity in
nats/s/Hz at high SNR becomes
\begin{align}    \label{eq:CumChSNR}
    \kappa_{n}
        \big|_{\snr ~\text{high}} =
        \tr\left\{
                \B{K}_{\left[1\right]}^{\left(n-1\right)}
                \left(0\right)
            \right\}
        +\delta_{1n}
        \cdot
        \left[
            \nXmin
            \ln\left(\frac{\snr}{\nTx}\right)
            +\ln\det\left(\CMmin\right)
        \right]
        +\sum_{\ell=1}^\nXmin
            \psi^{\left(n-1\right)}\left(\ell\right)
\end{align}
where
    $\delta_{ij}=
        \left\{
            \begin{smallmatrix}
                1, & i=j \\
                0, & i\ne j
            \end{smallmatrix}
        \right.$
is the Kronecker delta and
    $\psi^{\left(n\right)}\left(z\right) =
     d^{n+1} \ln\Gamma\left(z\right)/dz^{n+1}$
is the polygamma function.\footnote{
        For $z \in \mathbb{N}$, the digamma function
        $\psi^{\left(0\right)}\left(z\right)$,
        trigamma function $\psi^{\left(1\right)}\left(z\right)$,
        tetragamma function $\psi^{\left(2\right)}\left(z\right)$,
        and pentagamma function $\psi^{\left(3\right)}\left(z\right)$
        can be expressed as
        \begin{align*}
            \psi^{\left(0\right)}\left(z\right) =
                -\gamma
                +\sum_{n=1}^{z-1} \frac{1}{n},
            \quad
            \psi^{\left(1\right)}\left(z\right) =
                \frac{\pi^2}{6}
                -\sum_{n=1}^{z-1} \frac{1}{n^2},
            \quad
            \psi^{\left(2\right)}\left(z\right) =
                -2\zeta\left(3\right)
                +\sum_{n=1}^{z-1} \frac{2}{n^3},
            \quad
            \psi^{\left(3\right)}\left(z\right) =
                \frac{\pi^4}{15}
                -\sum_{n=1}^{z-1} \frac{6}{n^4}
        \end{align*}
        where $\gamma \approx 0.5772156649$ is the
        Euler-Mascheroni constant and $\zeta\left(3\right) \approx
        1.2020569$ is Ap\'{e}ry's constant.
        }
In particular, if $\nRx=\nTx$, \eqref{eq:CumChSNR} reduces to
\begin{align}    \label{eq:CumChSNRs}
    \kappa_{n}
        \big|_{\snr ~\text{high}} =
        \delta_{1n}
        \cdot
        \left[
                \nTx
                \ln\left(\frac{\snr}{\nTx}\right)
                +\ln\det\left(\TxCM \RxCM\right)
        \right]
        +\sum_{\ell=1}^{\nTx}
            \psi^{\left(n-1\right)}\left(\ell\right).
\end{align}

From \eqref{eq:CumChSNR} and \eqref{eq:CumChSNRs}, we have the
following observations.

\begin{itemize}

\item At high SNR, the mean capacity decreases by the amount of
$\ln\det\left(\CMmin\right)$ due to correlation at the side with
the smaller number of antennas and by the amount of
$\tr\left\{\B{K}_{[1]}(0)\right\}$ due to correlation at the side
with the larger number of antennas.

\item The variance, skewness, kurtosis, and other HOS of the
capacity depend only on correlation at the side with the larger
number of antennas and they converge to finite quantities
determined by \eqref{eq:CRM}--\eqref{eq:KurtC} and
\eqref{eq:CumChSNR}, as $\snr \to \infty$.

\item When $\nRx=\nTx$, fading correlation at any side does not
affect the variance and HOS in the high-SNR regime, while the mean
capacity decreases by the amount of $\ln\det\left(\TxCM
\RxCM\right)$. Moreover, it follows from \eqref{eq:SkewC},
\eqref{eq:KurtC}, and \eqref{eq:CumChSNRs} that
\begin{align*}
    -\frac{12\sqrt{6} \cdot \zeta\left(3\right)}{\pi^3}
    \leq
    \beta_{1}
        \big|_{\snr ~\text{high}}
    <0
    \quad
    \text{and}
    \quad
    0<
    \beta_{2}
        \big|_{\snr ~\text{high}}
    \leq
    2.4
\end{align*}
which imply that the capacity distribution has an asymmetric tail
extending out more to the left of its mean and is
\emph{leptokurtic} (i.e., more peaked than a Gaussian
distribution) in the high-SNR regime.

\item For a single-input single-output (SISO) case
($\nTx=\nRx=1$), we have
\begin{align*}
    m_{1} \big|_{\snr ~\text{high}}  =
        \ln\left(\snr\right)-\gamma,
    \quad
    \mu_{2} \big|_{\snr ~\text{high}}  =
        \frac{\pi^2}{6},
    \quad
    \beta_{1} \big|_{\snr ~\text{high}}  =
        -\frac{12\sqrt{6} \cdot \zeta\left(3\right)}{\pi^3},
    \quad
    \beta_{2} \big|_{\snr ~\text{high}}  =
        2.4
\end{align*}
which reveal that $-C$ at high SNR follows the \emph{extreme value
distribution} \cite{EHP00}.

\end{itemize}

\section{Numerical and Simulation Results}  \label{sec:Sec5}

To illustrate our analytical results, we consider the exponential
correlation model
\begin{align*}
    \ExpCM{n}{\rho} =
        \left[\rho^{|i-j|}\right]_{i,j=1,2,\ldots,n},
    \qquad
    \rho \in \left[0,1\right),
\end{align*}
as well as the multiple element transmit receive antennas (METRA)
model \cite{Ke02} in our numerical examples. The former model is
reasonable in the case of the equally-spaced linear array. The
latter model characterizes the correlation properties of MIMO
channels using a reduced set of physical parameters such as
antenna spacing, power angular spectrum, azimuth spread, and angle
of arrival. This model was validated based on measured data
collected in both picocell and microcell environments \cite{Ke02},
and also has been proposed recently for mobile broadband wireless
access (MBWA) MIMO channels \cite{C802.20-03/50}.

\subsection{Exponential Correlation Model}

\begin{figure}[t]
    \centerline{\includegraphics[width=0.7\textwidth]{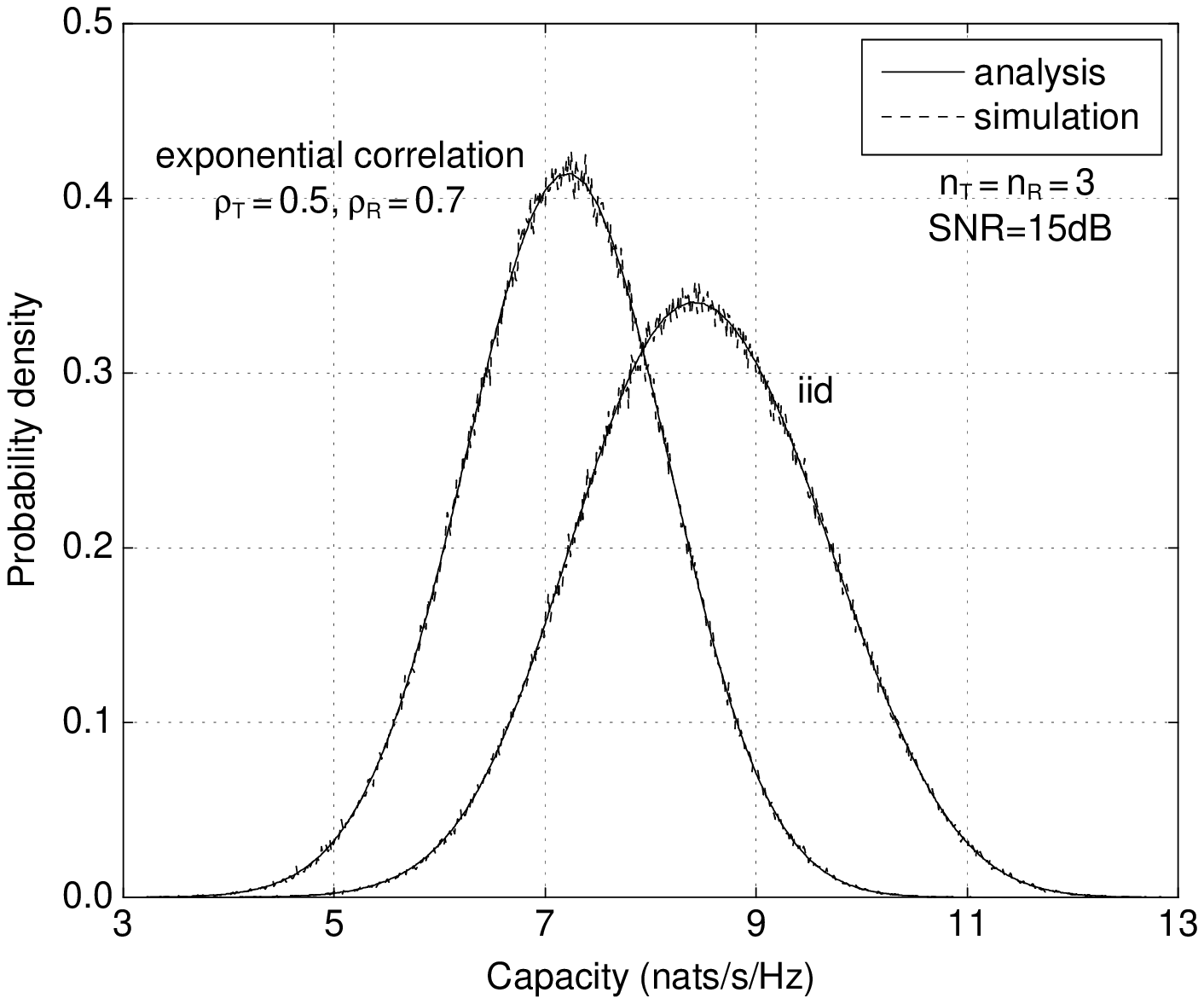}}
    \caption{
        PDF of the capacity for i.i.d. and exponentially
        correlated ($\TxCC =0.5$, $\RxCC =0.7$) MIMO channels.
        $\nTx=\nRx=3$ and $\snr=15$ dB.
    }
    \label{fig:Fig1}
\end{figure}

In all examples for exponential correlation, we set $\TxCM
=\ExpCM{\nTx}{\TxCC}$ and $\RxCM =\ExpCM{\nRx}{\RxCC}$. Fig.\
\ref{fig:Fig1} shows the PDF of $C$ for i.i.d.\ and exponentially
correlated ($\TxCC=0.5$, $\RxCC=0.7$) MIMO channels at $\snr=15$
dB when $\nTx=\nRx=3$. The analytical curves are plotted by using
\eqref{eq:CFCcor}, \eqref{eq:CFCiid}, and \cite[eq. (4.33)]{Sh04}.
We also compare our analytical results with the simulated PDF
obtained by generating $100\,000$ realizations of $\B{H}$. It can
be seen that analytical and simulated curves match exactly. The
figure also shows that the mass of the PDF is mostly above a
certain level due to the spatial multiplexing gain (for example, 4
nats/s/Hz for the exponentially correlated case and 5 nats/s/Hz
for the i.i.d.\ case).

\begin{figure}[t]
    \centerline{\includegraphics[width=0.7\textwidth]{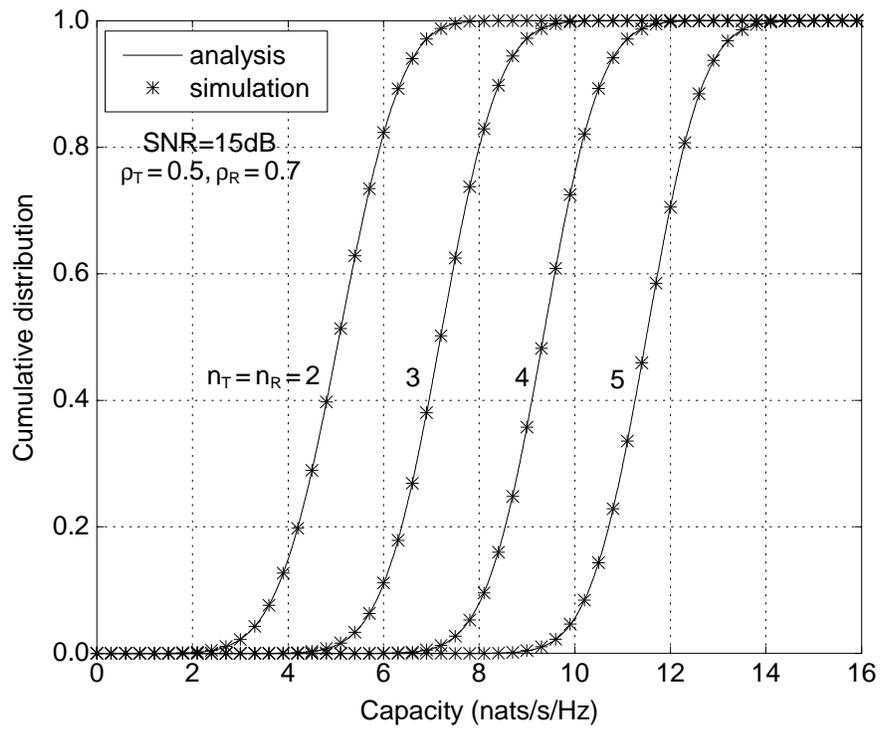}}
    \caption{
        CDF of the capacity for exponentially correlated MIMO
        channels with $\TxCC=0.5$ and $\RxCC=0.7$. $\nTx=\nRx=2,3,4,5$ and
        $\snr=15$ dB.
    }
    \label{fig:Fig2}
\end{figure}

\begin{figure}[t]
    \centerline{\includegraphics[width=0.7\textwidth]{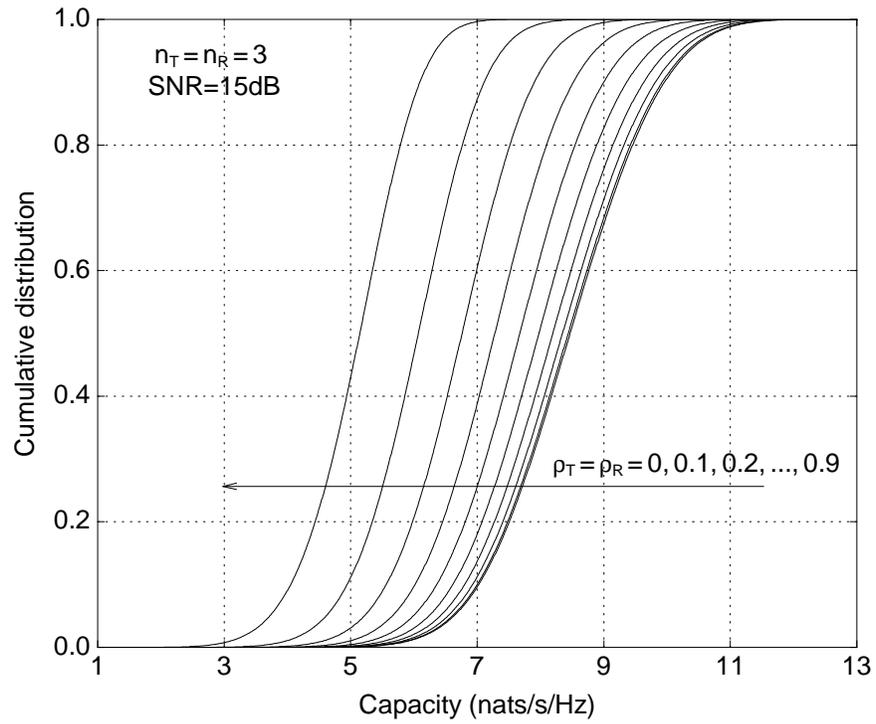}}
    \caption{
        CDF of the capacity for exponentially correlated MIMO
        channels with $\TxCC=\RxCC=0$ (i.i.d.), $0.1, 0.2, 0.3, 0.4, 0.5,
        0.6$, $0.7, 0.8$, and $0.9$. $\nTx=\nRx=3$ and $\snr=15$ dB.
    }
    \label{fig:Fig3}
\end{figure}

Fig.\ \ref{fig:Fig2} shows the CDF of $C$ for exponentially
correlated MIMO channels with $\TxCC=0.5$ and $\RxCC=0.7$ at
$\snr=15$ dB when $\nTx=\nRx=2$, $3$, $4$, and $5$. The analytical
curves are plotted by using \eqref{eq:CFCcor} and \cite[eq.
(4.34)]{Sh04}, and they agree exactly with the simulated ones. It
can be seen that the capacity increases linearly with the number
of antennas for the entire range of cumulative probability,
despite the presence of correlation. This can be attributed to the
spatial multiplexing gain achieved by increasing the number of
antennas at both sides. For example, the capacity at the
cumulative probability of $0.1$ (i.e., $10{\%}$ outage capacity)
is about $3.76$, $5.95$, $8.12$, and $10.30$ nats/s/Hz for
$\nTx=\nRx=2$, $3$, $4$, and $5$, respectively; we can gain
approximately $2.18$ nats/s/Hz of additional capacity for each
increase in the number of antennas at both the transmitter and the
receiver. Also, the spatial multiplexing gain of MIMO systems
guarantees a certain transmission rate at arbitrarily low outage
probability (e.g., $2.00$, $4.18$, $6.36$, and $8.54$ nats/s/Hz
for $\nTx=\nRx=2$, $3$, $4$, and $5$, respectively). Fig.\
\ref{fig:Fig3} illustrates the effect of exponential fading
correlation on the capacity distribution for the case of
$\nTx=\nRx=3$ and $\snr=15$ dB, where $\TxCC$ and $\RxCC$ range
from $0$ to $0.9$. It can be seen that the decrease in capacity
due to exponential correlation is negligible for a small amount of
correlation, but it becomes more significant as the correlation
coefficient increases. Moreover, the capacity reduction is more
pronounced at high cumulative (or outage) probability.

\begin{figure}[t]
    \centerline{\includegraphics[width=0.65\textwidth]{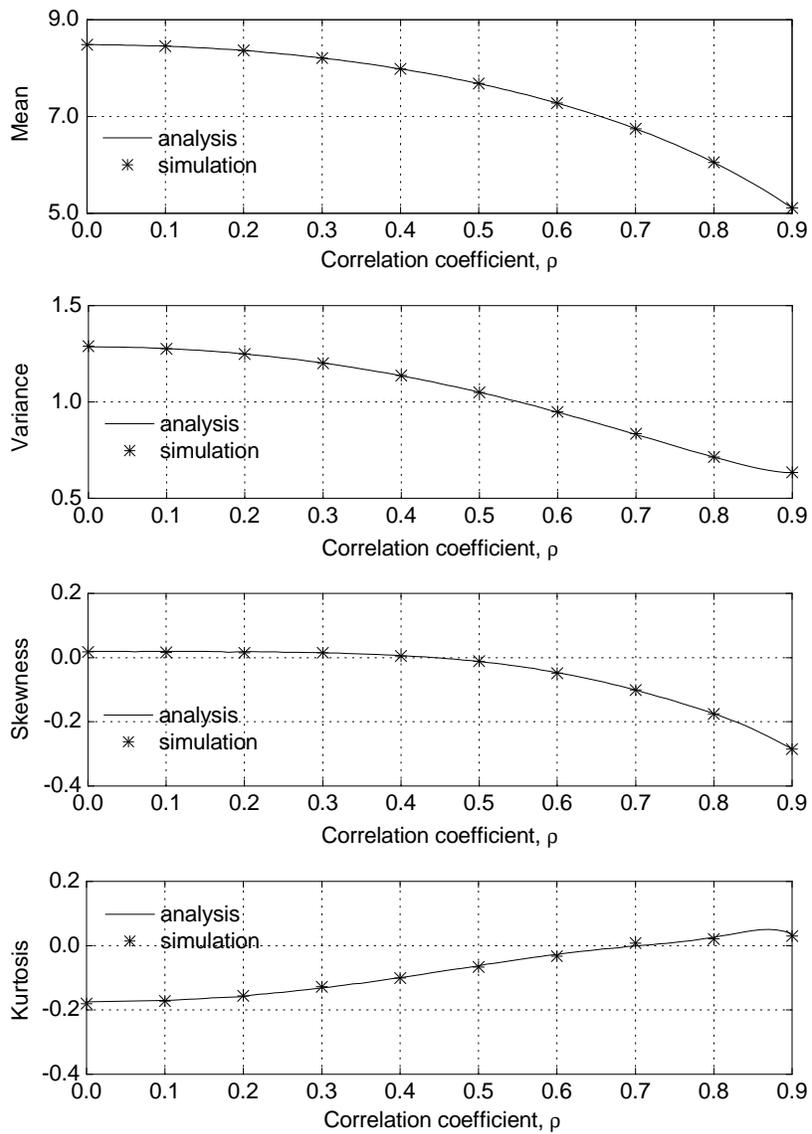}}
    \caption{
        Mean, variance, skewness, and kurtosis of the capacity
        (nats/s/Hz) as a function of correlation coefficient $\rho$ for
        exponentially correlated MIMO channels with $\TxCC =\RxCC=\rho$.
        $\nTx=\nRx=3$ and $\snr=15$ dB.
    }
    \label{fig:Fig4}
\end{figure}

\begin{figure}[t]
    \centerline{\includegraphics[width=0.65\textwidth]{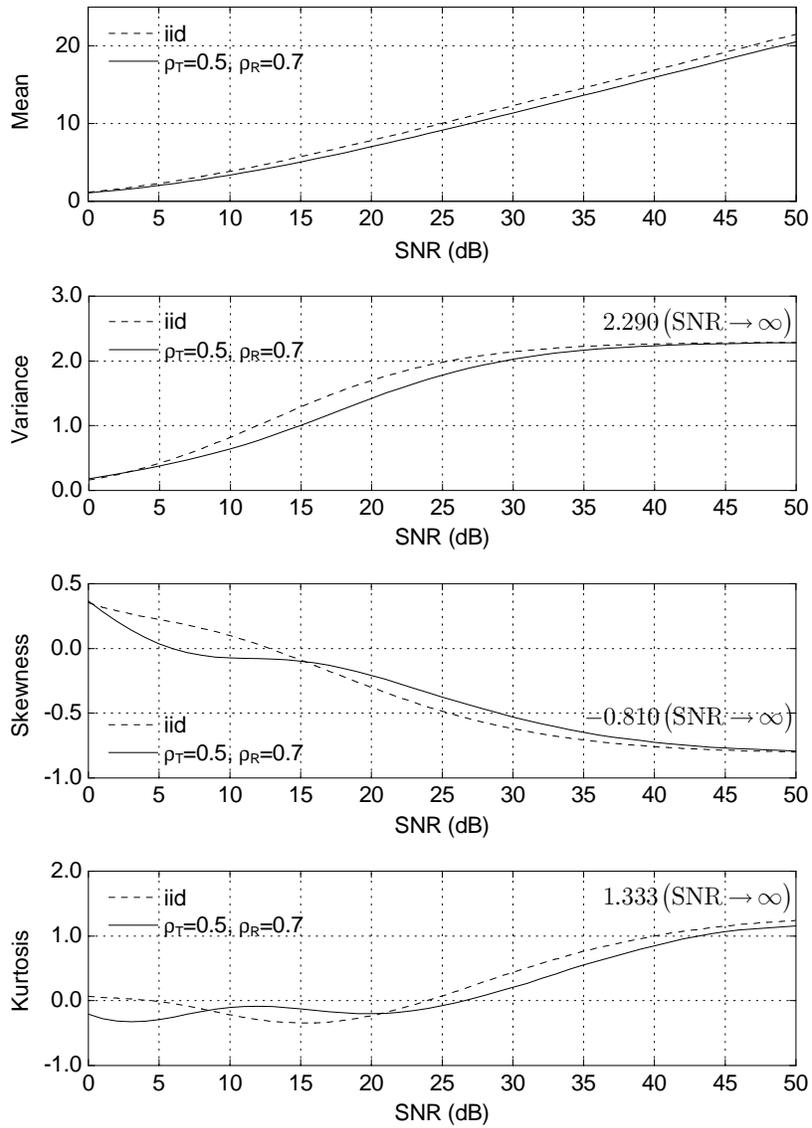}}
    \caption{
        Mean, variance, skewness, and kurtosis of the capacity
        (nats/s/Hz) versus SNR for i.i.d. and doubly correlated
        ($\TxCC=0.5$, $\RxCC=0.7$) MIMO channels. $\nTx=\nRx=2$.
    }
    \label{fig:Fig5}
\end{figure}

Fig.\ \ref{fig:Fig4} shows the analytical and simulated mean,
variance, skewness, and kurtosis of $C$ for exponentially
correlated MIMO channels as a function of correlation coefficient
$\rho$ for $\TxCC=\RxCC=\rho$, $\nTx=\nRx=3$, and $\snr=15$ dB.
Again, our analytical results are in excellent agreement with
Monte Carlo simulations that are carried out by generating
$100\,000$ realizations of $\B{H}$. Fig.\ \ref{fig:Fig5} shows the
mean, variance, skewness, and kurtosis of $C$ versus SNR for
i.i.d. and doubly correlated ($\TxCC=0.5$, $\RxCC=0.7$) MIMO
channels when $\nTx=\nRx=2$. It can be seen that as the SNR
increases, the variance, skewness, and kurtosis for both i.i.d.\
and doubly correlated cases converge to $2.290$, $-0.810$, and
$1.333$ according to \eqref{eq:SkewC}, \eqref{eq:KurtC}, and
\eqref{eq:CumChSNRs}, respectively, and the effect of fading
correlation on these statistics diminishes.

\subsection{METRA Correlation Model}

\begin{figure}[t]
    \centerline{\includegraphics[width=0.7\textwidth]{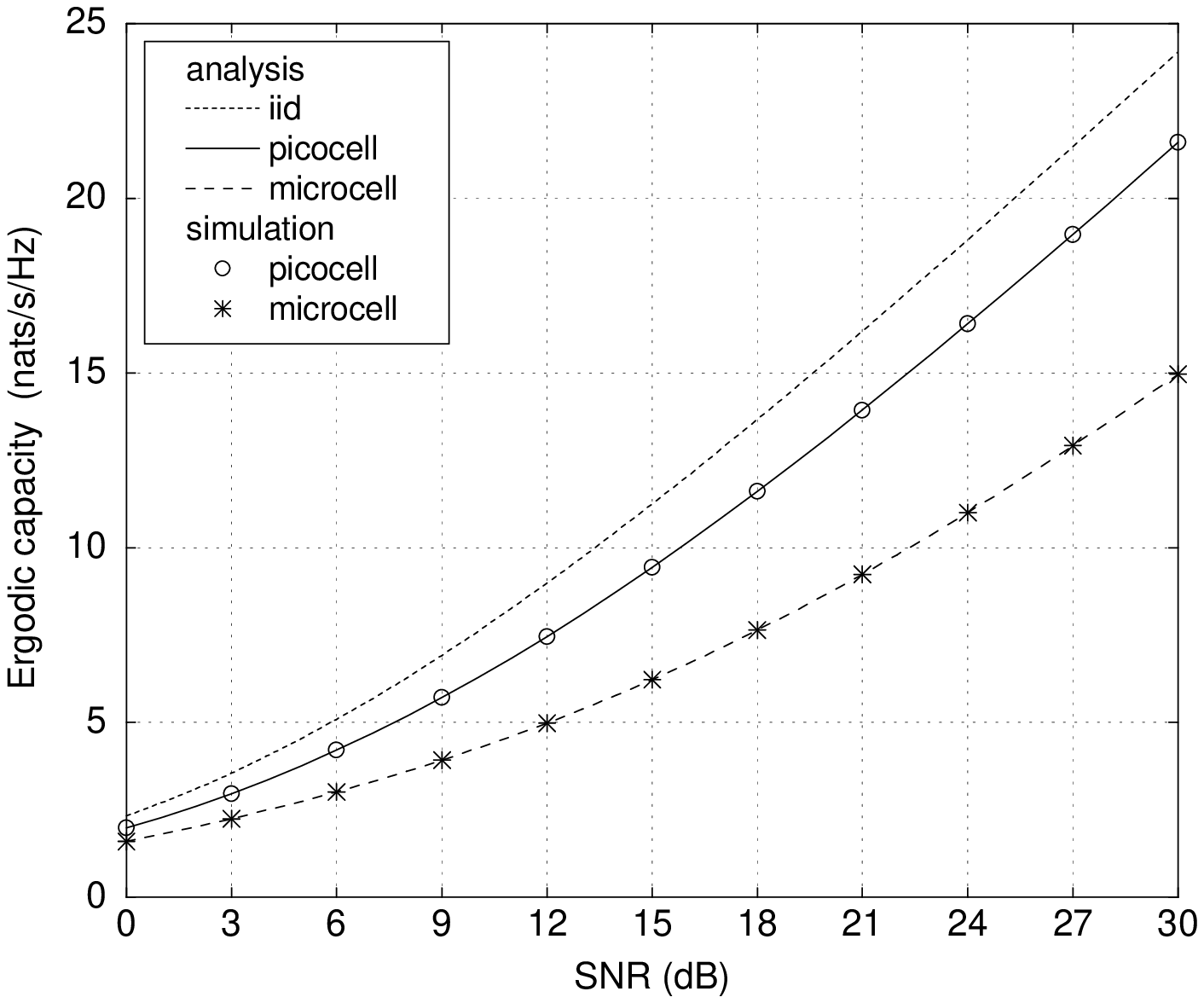}}
    \caption{
        Ergodic capacity of doubly correlated MIMO channels with
        $\nTx=\nRx=4$ and METRA correlation matrices given in
        \cite[p. 82]{Ke02} for picocell and microcell environments.
        For comparison, the ergodic capacity of the $4 \times 4$ i.i.d.
        MIMO channel is also plotted.
    }
    \label{fig:Fig6}
\end{figure}

\begin{figure}[t]
    \centerline{\includegraphics[width=0.7\textwidth]{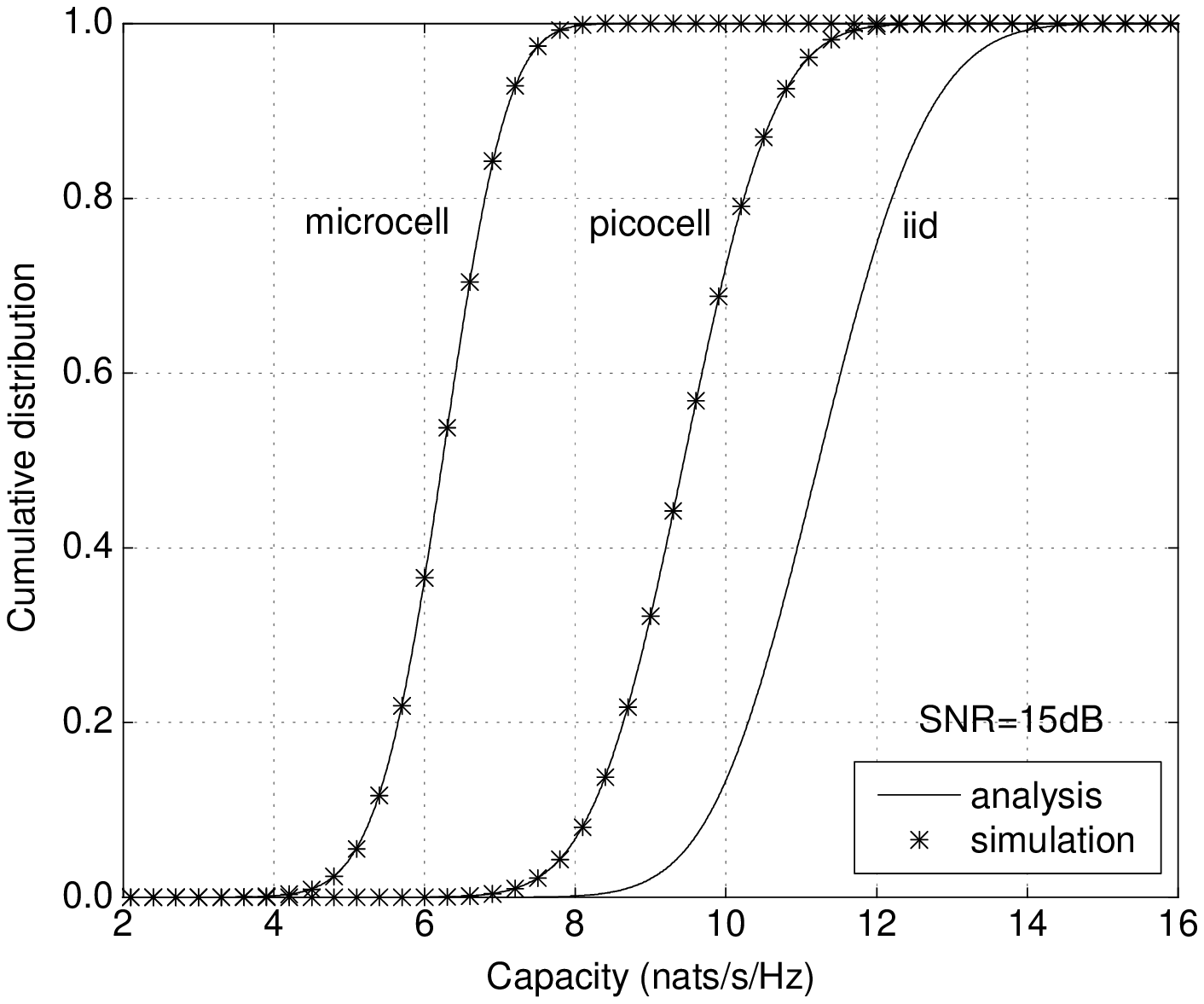}}
    \caption{
        CDF of the capacity for doubly correlated MIMO channels
        with $\nTx=\nRx=4$ and METRA correlation matrices given in
        \cite[p. 82]{Ke02} for picocell and microcell environments.
        For comparison, the CDF of capacity for the $4 \times 4$ i.i.d.
        MIMO channel is also plotted. $\snr=15$ dB.
    }
    \label{fig:Fig7}
\end{figure}

We now consider $4 \times 4$ MIMO channels with correlation
matrices obtained by the METRA model \cite{Ke02, C802.20-03/50}.
In all examples, it can be observed that our analytical results
agree exactly with Monte Carlo simulations. Fig.\ \ref{fig:Fig6}
shows the ergodic capacity for METRA correlation matrices, given
in \cite[p. 82]{Ke02}, for picocell and microcell environments.
The picocell example is a partially decorrelated scenario selected
from a small office environment, whereas the microcell case
corresponds to an environment where the receiver (base station) is
highly correlated (see \cite{Ke02} for details on the antenna
configurations and environment setups). We can see from Fig.\
\ref{fig:Fig6} that the ergodic capacity at $\snr=15$ dB is
$11.25$, $9.44$, and $6.22$ nats/s/Hz for i.i.d., picocell, and
microcell cases respectively. The reduction in ergodic capacity
due to spatial correlation is about $16\%$ for picocell and $45\%$
for microcell environments, respectively. Fig.\ \ref{fig:Fig7}
shows the CDF of the capacity at $\snr=15$ dB in the same
environments as in Fig. \ref{fig:Fig6}. The $10\%$ outage capacity
is $9.82$, $8.21$, and $5.33$ nats/s/Hz for i.i.d., picocell, and
microcell scenarios respectively. The reduction in $10\%$ outage
capacity due to spatial correlation is about $16\%$ for picocell
and $46\%$ for microcell environments, which is similar to the
amount of reduction in ergodic capacity.

\begin{figure}[t]
    \centerline{\includegraphics[width=0.7\textwidth]{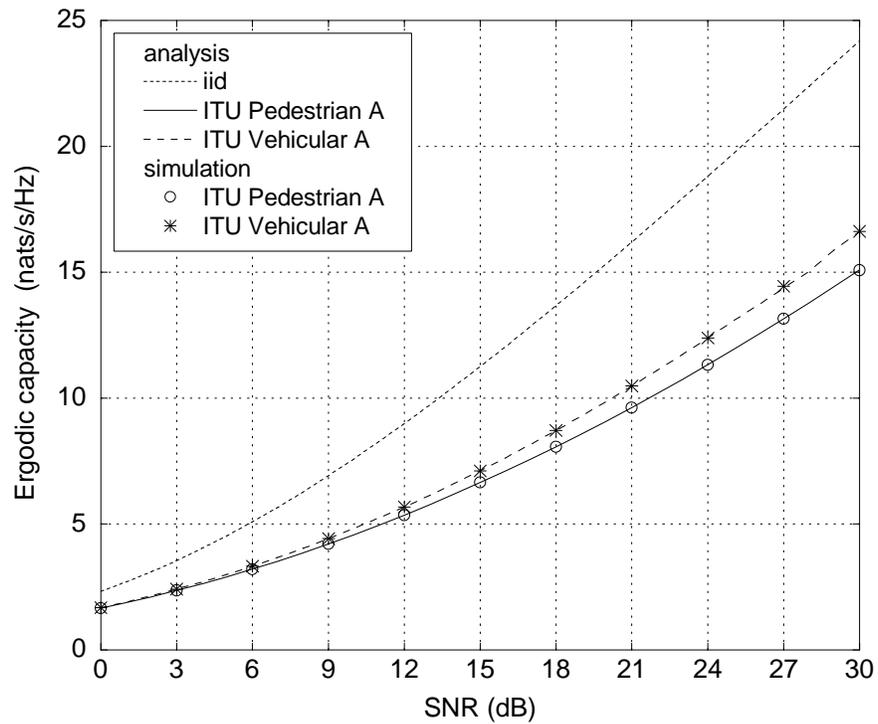}}
    \caption{
        Ergodic capacity of doubly correlated MIMO channels with
        $\nTx=\nRx=4$ and METRA correlation matrices given in
        \cite{C802.20-03/50} for macrocell ITU Pedestrian A and Vehicular
        A environments. For comparison, the ergodic capacity of the $4
        \times 4$ i.i.d. MIMO channel is also plotted.
    }
    \label{fig:Fig8}
\end{figure}

\begin{figure}[t]
    \centerline{\includegraphics[width=0.7\textwidth]{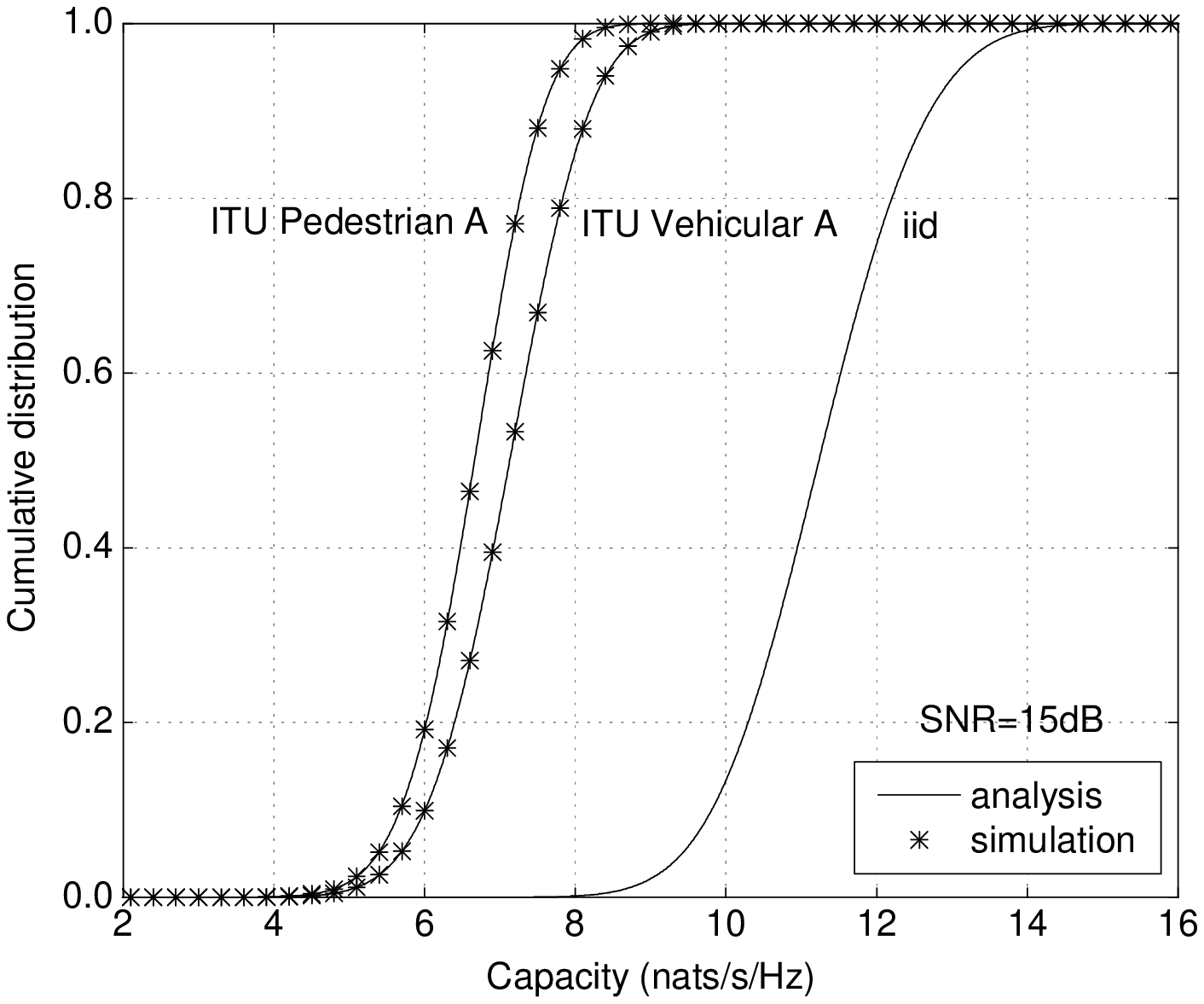}}
    \caption{
        CDF of the capacity for doubly correlated MIMO channels
        with $\nTx=\nRx=4$ and  METRA correlation matrices given in
        \cite{C802.20-03/50} for macrocell ITU Pedestrian A and Vehicular
        A environments. For comparison, the CDF of capacity for the $4
        \times 4$ i.i.d. MIMO channel is also plotted. $\snr=15$ dB.
    }
    \label{fig:Fig9}
\end{figure}

We next consider the METRA correlation matrices, given in
\cite{C802.20-03/50}, for $4\times 4$ MIMO channels in macrocell
Pedestrian A and Vehicular A environments of the international
telecommunication union (ITU) standard. For these correlation
matrices, the ergodic capacity is shown in Fig.\ \ref{fig:Fig8}
and the CDF of the capacity at $\snr=15$ dB is shown in Fig.\
\ref{fig:Fig9}. The ergodic capacity at $\snr=15$ dB is $6.64$
nats/s/Hz for the ITU Pedestrian A and $7.11$ nats/s/Hz for the
ITU Vehicular A, respectively (see Fig. \ref{fig:Fig8}). Also,
$10\%$ outage capacity is $5.67$ and $6.01$ nats/s/Hz for each
environment (see Fig. \ref{fig:Fig9}). The reduction in ergodic
capacity and $10\%$ outage capacity due to spatial correlation is
about $40\%$ in both environments.

\section{Conclusions} \label{sec:Sec6}

In this paper, we derived closed-form formulas for the exact
capacity statistics of Rayleigh-fading MIMO channels in the
presence of spatial fading correlation at both the transmitter and
the receiver. In particular, we derived the \emph{determinant}
representation for the characteristic function (Theorem
\ref{th:CFCcor}) and the \emph{trace} representations for the
cumulants (Theorems \ref{th:CumCiid} and \ref{th:CumCcor}) of MIMO
capacity as well as the mean, variance, skewness, and kurtosis
(Tables \ref{table:Table2} and \ref{table:Table3}). These results
are valid for arbitrary numbers of antennas, enabling us to
calculate both the ergodic capacity and the outage capacity
without any approximation and generalizing the previous results
for i.i.d.\ and one-sided correlated MIMO channels. We also showed
that in a high-SNR regime, the variance, skewness, kurtosis, and
other higher-order statistics of the capacity depend only on
correlation at the side with the larger number of antennas.
Moreover, when the antenna topology is symmetric (i.e.,
$\nTx=\nRx)$, these statistics are not affected by fading
correlation at any side and the capacity distribution has negative
skewness greater than or equal to $-12\sqrt{6} \cdot
\zeta\left(3\right)/{\pi^3}$, where $\zeta\left(3\right) \approx
1.2020569$ is Ap\'{e}ry's constant, and positive (excess) kurtosis
less than or equal to $2.4$. This implies that the capacity
distribution has an asymmetric tail extending out more to the left
of the ergodic capacity and a leptokurtic shape more peaked than a
Gaussian one. To illustrate our analytical results, we presented
numerical examples using the correlation model based on realistic
channel measurements as well as the classical exponential
correlation model. These examples showed that our analytical
results are in excellent agreement with Monte Carlo simulations
and that a considerable decrease in capacity, due to spatial
fading correlation, can be observed in realistic MIMO channels.

%
%
%

\section*{Appendix A: Integral Identities}

Let us define the integrals $\mathcal{G}_n \left(a,b,\xi\right)$
and $\mathcal{J}_{n,\ell} \left(a,b,\xi\right)$ as
\begin{align}    \label{eq:IntG}
    \mathcal{G}_n \left(a,b,\xi\right) \triangleq
        \int_0^\infty
            \left(1+ax\right)^{\xi-1}
            x^{n-1}
            e^{-x/b}
            dx,
        \quad
        a,b>0,
        ~n \in \mathbb{N},
        ~\xi \in \C
\end{align}
\begin{align}    \label{eq:IntJ}
    \mathcal{J}_{n,\ell} \left(a,b,\xi\right) \triangleq
        \frac{
            \partial^\ell
            \mathcal{G}_n \left(a,b,\xi\right)
            }
            {\partial \xi^\ell}
        = \int_0^\infty
            \left(1+ax\right)^{\xi-1}
            \ln^\ell \left(1+ax\right)
            x^{n-1}
            e^{-x/b}
            dx
\end{align}
which appear in deriving the analytical expressions for the
capacity statistics in Section \ref{sec:Sec3}.

From the integral representation of the confluent hypergeometric
function $\Psi \left(a,b;z\right)$ in \cite[eq. (9.211.4)]{GR00}
and the identity ${}_2F_0 \left(a,b;-z^{-1}\right)=z^a \Psi
\left(a,a-b+1;z\right)$, the integral
$\mathcal{G}_n\left(a,b,\xi\right)$ can be evaluated as
\begin{align}   \label{eq:GEval}
    \mathcal{G}_n \left(a,b,\xi\right)
        & = a^{-n}
            \left(n-1\right)!
            \Psi \left(n,n+\xi;\frac{1}{ab}\right)
        \nonumber \\
        & = b^n
            \left(n-1\right)! \;
            {}_2F_0 \left(n,-\xi+1;-ab\right)
\end{align}
where ${}_pF_q \left(a_1,a_2,\ldots,a_p;b_1,b_2,\ldots,b_q;z
\right)$ is the generalized hypergeometric function \cite[eq.
(9.14.1)]{GR00}. In particular, for $\xi \in \mathbb{N}$,
\eqref{eq:GEval} reduces to a finite sum of elementary functions
as
\begin{align}    \label{eq:Gs}
    \mathcal{G}_n \left(a,b,\xi\right)
        =   b^n
            \sum_{k=0}^{\xi-1}
                \binom{\xi-1}{k}
                \left(ab\right)^k
                \left(n+k-1\right)!.
\end{align}

Since the derivatives of the generalized hypergeometric function
with respect to its parameters are not known, in general, the
integral $\mathcal{J}_{n,\ell} \left(a,b,\xi\right)$ cannot be
evaluated directly from \eqref{eq:GEval}. However, $\mathcal{G}_n
\left(a,b,\xi\right)$ for $a,b>0$, $n \in \mathbb{N}$ and $\xi \in
\C$ can be expressed in an alternate form
\begin{align}    \label{eq:GEvala}
    \mathcal{G}_n \left(a,b,\xi\right)
        =   \frac{e^{1/\left(ab\right)}}{a^{n}}
            \sum_{k=0}^{n-1}
                \binom{n-1}{k}
                \left(-1\right)^{n-k-1}
                \left(ab\right)^{\xi+k}
                \Gamma\left(\xi+k,\frac{1}{ab}\right)
\end{align}
where $\Gamma\left(\alpha,z\right)=\int_z^\infty e^{-x} x^{\alpha
-1}dx$ is the complementary incomplete gamma function \cite[eq.
(8.350.2)]{GR00}. Then, using \eqref{eq:GEvala} and Leibniz's
identity \cite[p. 21]{GR00},
%
%
%
%
the integral $\mathcal{J}_{n,\ell} \left(a,b,\xi\right)$ defined
in \eqref{eq:IntJ} can be evaluated as
\begin{align}   \label{eq:JEval}
    \mathcal{J}_{n,\ell} \left(a,b,\xi\right)
        & = \frac{e^{1/\left(ab\right)}}{a^{n}}
            \sum_{k=0}^{n-1}
                \Biggl[
                    \left(-1\right)^{n-k-1}
                    \binom{n-1}{k}
                    \left(ab\right)^{\xi+k}
        \nonumber \\
        & \qquad \qquad \qquad \cdot
                \left.
                \sum_{i=0}^\ell
                    \left\{
                        \binom{\ell}{i}
                        \ln^{\ell-i}\left(ab\right)
                        \cdot
                        \left.
                            \left[
                                \frac{\partial^i}{\partial\alpha^i}
                                \Gamma\left(\alpha,\frac{1}{ab}\right)
                            \right]
                        \right|_{\alpha=\xi+k}
                    \right\}
                \right]
        \nonumber \\
        & = \frac{\ell! \; e^{1/\left(ab\right)}}{a^{n}}
            \sum_{k=0}^{n-1}
                \Biggl[
                    \left(-1\right)^{n-k-1}
                    \binom{n-1}{k}
                    \left(ab\right)^{\xi+k}
                    G_{\ell+1,\ell+2}^{\ell+2,0}
                        \Biggl(
                            \frac{1}{ab}
                            \Biggm|
                            \begin{subarray}{l}
                                \overbrace{1,1,\ldots,1}^{\ell+1 \text{ 1's}}
                                \\
                                \underbrace{0,0,\ldots,0}_{\ell+1 \text{ 0's}}, \;
                                \text{\normalsize $\xi+k$}
                            \end{subarray}
                        \Biggr)
                \Biggr]
\end{align}
where $G_{p,q}^{m,n} \left(\cdot\right)$ is the Meijer G-function
\cite[eq. (9.301)]{GR00}. In particular, for $\ell=\xi=1$,
\eqref{eq:JEval} reduces to
\begin{align}    \label{eq:Js}
    \mathcal{J}_{n,1} \left(a,b,1\right)
        =   b^n
            \left(n-1\right)!
            e^{1/\left(ab\right)}
            \sum_{k=0}^{n-1}
                \left(ab\right)^{-k}
                \Gamma\left(-k,\frac{1}{ab}\right).
\end{align}

\mySep


\section*{Appendix B: Proof of Theorem \ref{th:CFCcor}}

To proceed with the proof of Theorem \ref{th:CFCcor}, we begin by
evaluating an integral involving matrix determinants, which is a
continuous analogue of the well-known results in multivariate
analysis \cite{Kr76}. The next lemma adds a new identity to the
list of the generalized Cauchy-Binet formulas derived in
\cite[Appendix]{CWZ03IT}.

\begin{lemma}   \label{le:CB}

Suppose that $f_i $ and $g_j $, $i=1,2,\ldots ,m$, $j=1,2,\ldots
,n$, $m\leq n$, are arbitrary integrable functions over
$\mathfrak{D}$. Let $\B{F}\left(\varrho_1,\varrho_2
,\ldots,\varrho_m\right)$ and $\B{G}\left(\varrho_1,\varrho_2
,\ldots,\varrho_m\right)$ be $m\times m$ and $n\times n$ matrices
whose entries depend on $\varrho_1,\varrho_2 ,\ldots,\varrho_m$,
given by
\begin{gather}
    \left\{\B{F}\left(\varrho_1,\varrho_2
    ,\ldots,\varrho_m\right)\right\}_{i,j}
        =   f_j\left(\varrho_i\right),
        \qquad
        i,j=1,2,\ldots,m
    \\
    \left\{\B{G}\left(\varrho_1,\varrho_2
    ,\ldots,\varrho_m\right)\right\}_{i,j}
        =   \begin{cases}
                c_{i,j}, &  i=1,\ldots,n-m, ~j=1,\ldots,n
                \\[-3pt]
                g_j\left(\varrho_{i-n+m}\right), &   i=n-m+1,\ldots,n,
                ~j=1,\ldots,n
            \end{cases}
\end{gather}
where $c_{i,j}$ are scalar constants. Then,
\begin{align}    \label{eq:CB}
    \int_\mathfrak{D} \cdots \int_\mathfrak{D}
        \det\B{F}\left(\varrho_1,\varrho_2,\ldots,\varrho_m\right)
        \det\B{G}\left(\varrho_1,\varrho_2,\ldots,\varrho_m\right)
        \prod_{\ell=1}^m h\left(\varrho_\ell\right)
        d\varrho_1 d\varrho_2 \cdots d\varrho_m
    = m! \det\left(\B{\Phi}\right)
\end{align}
where $h\left(\cdot\right)$ is an arbitrary function and
$\B{\Phi}$ is the $n \times n$ matrix with $\left(i,j \right)$th
entry $\phi_{i,j}$ given by
\begin{align}
    \phi_{i,j}
        =   \begin{cases}
                c_{i,j}, &  i=1,\ldots,n-m, ~j=1,\ldots,n
                \\[-3pt]
                \int_\mathfrak{D}
                    f_{i-n+m}\left(\varrho\right)
                    g_j\left(\varrho\right)
                    h\left(\varrho\right)
                    d\varrho,
                    &   i=n-m+1,\ldots,n, ~j=1,\ldots,n.
            \end{cases}
\vspace*{3mm}%
\end{align}

\end{lemma}

\mySep

\begin{proof}
Let $\B{a}=\left(a_1,a_2,\ldots,a_n\right)$ and
$\B{b}=\left(b_1,b_2,\ldots,b_m\right)$ be the permutations of
integers $1,2,\ldots,n$ and $1,2,\ldots,m$, respectively. Then,
the integration of the left-hand side of \eqref{eq:CB}, denoted by
$\mathcal{I}$, becomes
%
%
\begin{align}
    \mathcal{I}
    =
        \sum_{\B{b}}
            \sgn\left(\B{b}\right)
        \sum_{\B{a}}
            \sgn\left(\B{a}\right)
            \prod_{i=1}^{n-m} c_{i,a_i}
            \prod_{j=1}^m \phi_{n-m+b_j,a_{n-m+j}}
        \, ,
\end{align}
where $\sgn\left(\cdot\right)$ denotes the sign of the
permutation. Note that the sequence
$1,2,\ldots,n-m,b_1+n-m,b_2+n-m,\ldots,b_m+n-m$ is a permutation
of $1,2,\ldots,n$ with the sign equal to $\sgn\left(\B{b}\right)$.
Therefore, using \cite[eq.~(38)]{CWZ03IT}, we get
\begin{align}
    \mathcal{I}
    =
        \sum_{\B{b}}
            \det\left(\B{\Phi}\right)
    =
        m! \det\left(\B{\Phi}\right).
\end{align}
\end{proof}

\mySep

\textit{Proof of Theorem \ref{th:CFCcor}:} The CF\ of $C$ can be
written as
\begin{align}    \label{eq:CCp0}
    \CF{C}{\jmath\omega}
        =   \int_{\MHH=\MHH^\dag>0}
                \det\left(\B{I}_\nXmin+\bar{\snr}\MHH\right)^{\jmath\omega}
                \PDF{\B{\Theta}}{\MHH}
                d\MHH
\end{align}
where $\bar{\snr}=\snr/\nTx$ and from \eqref{eq:CQM}, the PDF of
$\MHH \sim \CQM{\nXmin}{\nXmax}{\B{I}_\nXmax}{\CMmin}{\CMmax}$ is
given by
\begin{align}    \label{eq:PDFth}
    \PDF{\MHH}{\MHH}
        =   \frac{1}{\tilde{\Gamma}_\nXmin\left(\nXmax\right)}
            \det\left(\CMmin\right)^{-\nXmax}
            \det\left(\CMmax\right)^{-\nXmin}
            \det\left(\MHH\right)^{\nXmax-\nXmin}
            {}_0\tilde{F}_0^{\left(\nXmax\right)}
                \left(-\CMmin^{-1}\MHH,\CMmax^{-1}\right).
\end{align}
The typical approach for the evaluation of the integral in
\eqref{eq:CCp0} is to perform eigenvalue decomposition using the
unitary transformation of $\MHH$ and to exploit the knowledge of
the joint eigenvalue distribution of $\MHH$. However, the
correlation matrix $\CMmin$ in the argument of the hypergeometric
function in \eqref{eq:PDFth} prevents the removal of the unitary
matrix from its arguments after the eigenvalue decomposition,
which makes it difficult to directly use the joint eigenvalue
distribution of $\MHH$. We alleviate this difficulty by performing
two successive transformations as follows. The first
transformation is given by $\B{Z}=\CMmin^{-1/2}\MHH\CMmin^{-1/2}$
with Jacobian $d\MHH=\det\left(\CMmin\right)^\nXmin d\B{Z}$.
Using the fact that
\begin{align}
    {}_0\tilde{F}_0^{\left(\nXmax\right)}
        \left(-\CMmin^{-1}\MHH,\CMmax^{-1}\right)
        =   {}_0\tilde{F}_0^{\left(\nXmax\right)}
                \left(
                    -\CMmin^{-1/2}\MHH\CMmin^{-1/2},\CMmax^{-1}
                \right),
\end{align}
we have
\begin{align}    \label{eq:CCp1}
    \CF{C}{\jmath\omega}
        =   \frac{\det\left(\CMmax\right)^{-\nXmin}}
                 {\tilde{\Gamma}_\nXmin\left(\nXmax\right)}
            \int_{\B{Z}=\B{Z}^\dag >0}
         &       \det\left(\B{I}_\nXmin+\bar{\snr}\CMmin \B{Z}\right)^{\jmath\omega}
        \nonumber \\
        &
                \cdot \det\left(\B{Z}\right)^{\nXmax-\nXmin}
                {}_0\tilde{F}_0^{\left(\nXmax\right)}
                    \left(-\B{Z},\CMmax^{-1}\right)
                d\B{Z}.
\end{align}

Let us denote a unitary manifold of $\nXmin \times \nXmin$ unitary
matrices with real diagonal elements by
$\tilde{U}\left(\nXmin\right)$. Since $\B{Z}$ is Hermitian, there
exists $\B{U} \in \tilde{U}\left(\nXmin\right)$ such that
$\B{Z}=\B{UDU}^\dag$ and $\B{D}=\mathrm{diag}\left(z_1,z_2,\ldots
,z_\nXmin\right)$ where $0<z_1 \leq z_2\leq \cdots \leq z_\nXmin$
are ordered eigenvalues of $\B{Z}$. We then make the second
transformation $\B{Z}=\B{UDU}^\dag$ with Jacobian
$d\B{Z}=\prod_{1\leq i<j \leq \nXmin} \left(z_j -z_i\right)^2
d\B{U}d\B{D}$ \cite[Theorem 3.1]{Ed89}, \cite[Theorem 4.4]{Ma97},
yielding
%
%
%
%
%
%
\begin{align}   \label{eq:CCp2}
    \CF{C}{\jmath\omega}
        =   \frac{\det\left(\CMmax\right)^{-\nXmin}}
                 {\tilde{\Gamma}_\nXmin\left(\nXmax\right)}
            \int_{\B{D}}
            &
            \int_{\B{U}\in\tilde{U}{\left(\nXmin\right)}}
                \det\left(\B{I}_\nXmin+\bar{\snr} \CMmin \B{UDU}^\dag \right)^{\jmath\omega}
                \det\left(\B{D}\right)^{\nXmax-\nXmin}
            \nonumber \\
            & \qquad \quad \cdot
            \prod_{1\leq i<j\leq \nXmin}\left(z_j -z_i\right)^2
            {}^{}_0\tilde{F}_0^{\left(\nXmax\right)} \left(-\B{D},\CMmax^{-1}\right)
            d\B{U}d\B{D}
\end{align}
where we have used the fact that the hypergeometric function with
matrix arguments is invariant under unitary transformations of its
arguments.\footnote{
        Note that $\prod_{1 \leq i<j
        \leq \nXmin} \left( z_j - z_i \right)$ is the $\nXmin \times
        \nXmin$ Vandermonde determinant of $z_1,z_2,\ldots,z_\nXmin$,
        i.e.,
        \begin{align*}
            \prod_{1 \leq i<j \leq \nXmin} \left( z_j - z_i \right)
                =   \det\left(\left[z_j^{i-1}\right]\right).
        \end{align*}
        }
It is now apparent that the above two transformations enable us to
remove the dependence of $\B{U}$ on the hypergeometric function.
Recall that the total volume of $\tilde{U}\left(\nXmin\right)$ is
\cite[Corollary 4.3.1]{Ma97}
\begin{align}    \label{eq:UV}
    \int_{\B{U} \in \tilde{U}\left(\nXmin\right)} d\B{U}
        =   \frac{\pi^{\nXmin\left(\nXmin-1\right)}}
                 {\tilde{\Gamma}_\nXmin \left(\nXmin\right)}.
\end{align}
We can now carry out the integration with respect to $\B{U}$ using
\cite[eq. (6.1.19)]{Ma97} and \cite[eq. (52)]{Kh66} as
\begin{align}    \label{eq:CCp3}
    \int_{\B{U} \in \tilde{U}\left(\nXmin\right)}
        \det\left(\B{I}_\nXmin + \bar{\snr} \CMmin \B{UDU}^\dag\right)^{\jmath\omega}
    d\B{U}
    =   \frac{\pi^{\nXmin\left(\nXmin-1\right)}}
             {\tilde{\Gamma}_\nXmin\left(\nXmin\right)}
        {}_1\tilde{F}_0^{\left(\nXmin\right)}
            \left(-\jmath \omega ; \B{D}, -\bar{\snr} \CMmin \right).
\end{align}
Substituting \eqref{eq:CCp3} into \eqref{eq:CCp2} gives
\begin{align}   \label{eq:CCp4}
    \CF{C}{\jmath\omega}
        = &  \frac{\pi^{\nXmin\left(\nXmin-1\right)}
                    \det\left(\CMmax\right)^{-\nXmin}}
                 {\tilde{\Gamma}_\nXmin \left(\nXmax\right)
                    \tilde{\Gamma}_\nXmin \left(\nXmin\right)}
            \int_{0<z_1\leq \cdots \leq z_\nXmin<\infty}
                \prod_{\ell=1}^\nXmin z_\ell^{\nXmax-\nXmin}
                \prod_{1\leq i<j\leq \nXmin} \left(z_j -z_i \right)^2
            \nonumber \\
            & \hspace{3cm}
            \cdot
            {}_1\tilde{F}_0^{\left(\nXmin\right)}
                \left(-\jmath \omega ; \B{D}, -\bar{\snr} \CMmin \right)
            {}_0\tilde{F}_0^{\left(\nXmax\right)}
                \left(-\B{D}, \CMmax^{-1} \right)
            dz_1 dz_2 \cdots dz_\nXmin.
\end{align}
Using the results in \cite[Lemma 3]{Kh70} and \cite{GS00}, the
hypergeometric functions with matrix arguments in the integrand of
\eqref{eq:CCp4} can be expressed in terms of determinants as

\begin{gather}    \label{eq:F10}
    {}_1\tilde{F}_0^{\left(\nXmin\right)}
        \left(-\jmath \omega ; \B{D}, -\bar{\snr} \CMmin \right)
       =   \frac{
                    \left(-\bar{\snr}\pi\right)^{-\nXmin\left(\nXmin-1\right)/2}
                    \tilde{\Gamma}_\nXmin \left(\nXmin\right)
                    \det\left[
                                \left(1+\bar{\snr}\Emin{j} z_i\right)^{\jmath\omega+\nXmin-1}
                        \right]_{i,j=1,2,\ldots,\nXmin}
                }
                {
                    \prod_{\ell=1}^\nXmin \left(-\jmath\omega-\nXmin+1\right)_{\ell-1}
                    \prod_{1\leq i<j\leq \nXmin} \left(z_j -z_i\right)
                    \left(\Emin{j}-\Emin{i}\right)
                }
    \\[10pt]
    \label{eq:F00}
    {}_0\tilde{F}_0^{\left(\nXmax\right)}
        \left(-\B{D}, \CMmax^{-1} \right)
        =   \frac{
                    \pi^{-\nXmin\left(\nXmin-1\right)/2}
                    \tilde{\Gamma}_\nXmin \left(\nXmax\right)
                    \det\left(\CMmax\right)^\nXmin
                    \det\left(\B{\Xi}\right)
                }
                {
                    \prod_{\ell=1}^\nXmin z_\ell^{\nXmax-\nXmin}
                    \prod_{1\leq i<j\leq \nXmin} \left(z_j -z_i \right)
                    \prod_{1\leq i<j\leq \nXmax} \left(\Emax{j}-\Emax{i}\right)
                }
\end{gather}
\mySep

\noindent where $\left(\alpha\right)_n=\alpha
\left(\alpha+1\right)\cdots \left(\alpha+n-1\right)$,
$\left(\alpha\right)_0=1$, is the Pochhammer symbol and $\B{\Xi}$
is the $\nXmax \times \nXmax$ matrix whose $\left(i,j\right)$th
entry is given by
\begin{align}    \label{eq:Mdelta}
    \left\{\B{\Xi}\right\}_{i,j}
        =   \begin{cases}
                \Emax{j}^{i-1}, &   i=1,\ldots,\nXmax-\nXmin,~j=1,\ldots,\nXmax
                \\
                \Emax{j}^{\nXmax-\nXmin-1}
                \exp\left(-\frac{z_{i-\nXmax+\nXmin}}{\Emax{j}}\right),
                &
                i=\nXmax-\nXmin+1,\ldots,\nXmax,~j=1,\ldots,\nXmax.
            \end{cases}
\end{align}
Combining \eqref{eq:CCp4}--\eqref{eq:Mdelta} together with
\begin{align}
    \left(-1\right)^{\nXmin\left(\nXmin-1\right)/2}
    \prod_{\ell=1}^\nXmin \left(-\jmath\omega -\nXmin+1 \right)_{\ell-1}
    =   \prod_{\ell=1}^{\nXmin-1} \left(\jmath\omega+\ell\right)^\ell,
\end{align}
we get
\begin{align}   \label{eq:CCp5}
    \CF{C}{\jmath\omega}
    & = \frac{\Upsilon_\nXmin\left(\jmath\omega\right)}{\Kc}
        \int_{0<z_1\leq \cdots \leq z_\nXmin<\infty}
            \det\left[
                    \left(1+\bar{\snr}\Emin{j} z_i\right)^{\jmath\omega+\nXmin-1}
                \right]_{i,j=1,2,\ldots,\nXmin}
            \det\left(\B{\Xi}\right)
            dz_1 \cdots dz_\nXmin
    \nonumber \\
    & = \frac{\Upsilon_\nXmin\left(\jmath\omega\right)}{\nXmin!\Kc}
        \int_0^\infty \cdots \int_0^\infty
            \det\left[
                    \left(1+\bar{\snr}\Emin{j} z_i\right)^{\jmath\omega+\nXmin-1}
                \right]_{i,j=1,2,\ldots,\nXmin}
            \det\left(\B{\Xi}\right)
            dz_1 \cdots dz_\nXmin
\end{align}
where the last equality follows from the fact that the integrand
is symmetric in $z_1,z_2,\ldots,z_\nXmin$. Finally, applying the
integral-type Cauchy-Binet formula in Lemma \ref{le:CB} to
\eqref{eq:CCp5} and using the identity \eqref{eq:GEval} complete
the proof.

\section*{Acknowledgment}

The authors would like to thank the Associate Editor and the
anonymous reviewers for their helpful suggestions and comments.

\bibliographystyle{IEEEtran}
\bibliography{IEEEabrv,Capacity-DC-TW04-850-R2}

\clearpage

\renewcommand{\arraystretch}{1.5}

\begin{table}[t]
    \caption{
        Some matrices involved in the analytical expressions for
        the capacity statistics (see Appendix A 
        for details on
        the evaluation of integrals). Denote $~\bar{\snr}=\snr/\nTx$.
    }
    \label{table:Table1}
    \vspace*{-2mm}
    \begin{center}
    \begin{tabular}{c|c|l}
    \hline
    \raisebox{1.5pt}[0mm]
    {~~Notation~~}
    &
    \raisebox{1.5pt}[0mm]
    {Dimension}
    &
    \raisebox{1.5pt}[0mm]
    {$~\left(i,j\right)$th entry
    (where $\imath=i-\nXmax+\nXmin$ and $n\in \mathbb{N}$)}
    \\
    \hline \hline
    $\MC\left(\nu\right)$
    &
    $\nXmax \times \nXmax$
    &
    $\begin{array}{l}
        \left\{\MC\left(\nu\right)\right\}_{
            i=1,2,\dots,\nXmax-\nXmin,
            j=1,2,\dots,\nXmax
        }
        = \Emax{j}^{i-1}
        \\
        \left\{\MC\left(\nu\right)\right\}_{
            i=\nXmax-\nXmin+1,\nXmax-\nXmin+2,\dots,\nXmax,
            j=1,2,\dots,\nXmax
        }
        \\
        \quad
        = \displaystyle{
            \Emax{j}^{\nXmax-\nXmin-1}
            \int_0^\infty
                \left(1+\bar{\snr}\Emin{\imath}z\right)^{\nu+\nXmin-1}
                e^{-z/\Emax{j}}
                dz
            }
        \\
        \quad
        = \Emax{j}^{\nXmax-\nXmin} \;
            {}_2F_0\left(1,-\nu-\nXmin+1;-\bar{\snr}\Emin{\imath}\Emax{j}\right)
    \end{array}$
    \\
    \hline
    $\MC^{\left(n\right)}\left(\nu\right)$
    &
    $\nXmax \times \nXmax$
    &
    $\begin{array}{l}
        \left\{\MC^{\left(n\right)}\left(\nu\right)\right\}_{
            i=1,2,\dots,\nXmax-\nXmin,
            j=1,2,\dots,\nXmax
        }
        = 0
        \\
        \left\{\MC^{\left(n\right)}\left(\nu\right)\right\}_{
            i=\nXmax-\nXmin+1,\nXmax-\nXmin+2,\dots,\nXmax,
            j=1,2,\dots,\nXmax
        }
        \\
        \quad
        = \displaystyle{
            \Emax{j}^{\nXmax-\nXmin-1}
            \int_0^\infty
                \left(1+\bar{\snr}\Emin{\imath}z\right)^{\nu+\nXmin-1}
                \ln^n \left(1+\bar{\snr}\Emin{\imath}z\right)
                e^{-z/\Emax{j}}
                dz
            }
        \\
        \quad
        =   n!
            \left(\bar{\snr}\Emin{\imath}\right)^{\nu+\nXmin-1}
            \Emax{j}^{\nu+\nXmax-1}
            \exp\left(\frac{1}{\bar{\snr}\Emin{\imath}\Emax{j}}\right)
        \\
        \qquad \times
        G_{n+1,n+2}^{n+2,0}
            \left(
                \left.
                    \frac{1}{\bar{\snr}\Emin{\imath}\Emax{j}}
                \right|
                \begin{subarray}{l}
                    1,1,\ldots,1
                    \\
                    0,0,\ldots,0,\nu+\nXmin
                \end{subarray}
            \right)
    \end{array}$
    \\
    \hline
    $\MI\left(\nu\right)$
    &
    $\nXmin \times \nXmin$
    &
    $\begin{array}{l}
        \left\{\MI\left(\nu\right)\right\}_{
            i,j=1,2,\ldots,\nXmin
        }
        \\
        \quad
        =   \displaystyle{
                \int_0^\infty
                    \left(1+\bar{\snr}z\right)^{\nu}
                    z^{\nXmax-\nXmin+i+j-2}
                    e^{-z}
                    dz
            }
        \\
        \quad
        =   \left(\nXmax-\nXmin+i+j-2\right)! \;
            {}_2F_0\left(\nXmax-\nXmin+i+j-1,-\nu;-\bar{\snr}\right)
    \end{array}$
    \\
    \hline
    $\MI^{\left(n\right)}\left(\nu\right)$
    &
    $\nXmin \times \nXmin$
    &
    $\begin{array}{l}
        \left\{\MI^{\left(n\right)}\left(\nu\right)\right\}_{
            i,j=1,2,\ldots,\nXmin
        }
        \\
        \quad
        =   \displaystyle{
                \int_0^\infty
                    \left(1+\bar{\snr}z\right)^{\nu}
                    \ln^n \left(1+\bar{\snr}z\right)
                    z^{\nXmax-\nXmin+i+j-2}
                    e^{-z}
                    dz
            }
        \\
        \quad
        =   \displaystyle{
                \frac{
                        n!
                        e^{1/\bar{\snr}}
                    }
                    {\bar{\snr}^{\nXmax-\nXmin+i+j-1}}
                \sum_{k=0}^{\nXmax-\nXmin+i+j-2}
                    \Biggl[
                        \left(-1\right)^{\nXmax-\nXmin+i+j-k-2}
            }
        \\
        \qquad \times
            \binom{\nXmax-\nXmin+i+j-2}{k}
            \bar{\snr}^{\nu+k+1}
            G_{n+1,n+2}^{n+2,0}
                \left(
                    \left.
                        \frac{1}{\bar{\snr}}
                    \right|
                    \begin{subarray}{l}
                        1,1,\ldots,1
                        \\
                        0,0,\ldots,0,\nu+k+1
                    \end{subarray}
                \right)
            \Biggr]
    \end{array}$
    \\
    \hline
    \end{tabular}
    \end{center}
\end{table}

\renewcommand{\arraystretch}{1}

\clearpage

\renewcommand{\arraystretch}{1.2}

\begin{table}[t]
    \vspace*{2cm}
    \caption{
        Mean, variance, skewness, and kurtosis of the capacity in
        nats/s/Hz for i.i.d. MIMO channels, $\B{H} \sim
        \CGM{\nRx}{\nTx}{\B{0}}{\B{I}_\nRx}{\B{I}_\nTx}.~$  Denote
        $~\MI_n = \MI_{[n]} \left(0\right)$.
        }
    \label{table:Table2}
    \vspace*{-1mm}
    \begin{center}
    \begin{tabular}{p{0.01mm}ll p{0.01mm}}
\hline
&   $m_{1} $ (mean)    &
                                        $\tr\left(\MI_1\right)$ &
                                        \\[-0.7mm]
&   $\mu_{2}$ (variance)        &
                                        $\tr\left(\MI_2-\MI_1^2\right)$
                                        &
                                        \\[1.7mm]
&   $\beta_{1}$ (skewness)       &
                                        $\displaystyle{\frac{\tr\left(2\,\MI_1^3-3\,\MI_1 \MI_2 +\MI_3\right)}
                                        {\left[\tr\left(\MI_2-\MI_1^2\right)\right]^{3/2}}}$
                                        &
                                        \\[4.2mm]
&   $\beta_{2}$ (kurtosis)       &
                                        $\displaystyle{\frac{\tr\left(-6\,\MI_1^4+12\,\MI_1^2 \,\MI_2 -3\,\MI_2^2
                                        -4\,\MI_1 \MI_3 +\MI_4
                                        \right)}{\left[\tr\left(\MI_2-\MI_1^2\right)\right]^{2}}}
                                        $ &
                                        \\[3.5mm]
\hline
\end{tabular}
\end{center}
\end{table}

\begin{table}[t]
    \caption{
        Mean, variance, skewness, and kurtosis of the capacity in
        nats/s/Hz for doubly correlated MIMO channels, $~\B{H} \sim
        \CGM{\nRx}{\nTx}{\B{0}}{\RxCM}{\TxCM}.~$  Denote
        $~\MC_n = \MC_{[n]} \left(0\right)$.}
    \label{table:Table3}
    \vspace*{-1mm}
    \begin{center}
    \begin{tabular}{p{0.01mm}ll p{0.01mm}}
\hline
&   $m_{1} $ (mean)    &
                                        $\tr\left(\MC_1\right)-\left(\nXmin-1\right)$ &
                                        \\[0mm]
&   $\mu_{2}$ (variance)        &
                                        $\displaystyle{\tr\left(\MC_2-\MC_1^2\right)+\sum_{\ell=1}^{\nXmin-1} \ell^{-1}}$
                                        &
                                        \\[3.8mm]
&   $\beta_{1}$ (skewness)       &
                                        $\displaystyle{\frac{\tr\left(2\MC_1^3-3\MC_1 \MC_2
                                        +\MC_3\right)-\sum_{\ell=1}^{\nXmin-1} 2\ell^{-2}}
                                        {\left[\tr\left(\MC_2-\MC_1^2\right)+\sum_{\ell=1}^{\nXmin-1}
                                        \ell^{-1}\right]^{3/2}}} $
                                        &
                                        \\[4.5mm]
&   $\beta_{2}$ (kurtosis)       &
                                        $\displaystyle{\frac{\tr\left(-6\MC_1^4+12\MC_1^2 \MC_2 -3\MC_2^2
                                        -4\MC_1 \MC_3 +\MC_4 \right)+\sum_{\ell=1}^{\nXmin-1} 6\ell^{-3}}
                                        {\left[\tr\left(\MC_2-\MC_1^2\right)+\sum_{\ell=1}^{\nXmin-1}
                                        \ell^{-1}\right]^{2}}} $  &
                                        \\[4mm]
\hline
\end{tabular}
\end{center}
\vspace*{2cm}
\end{table}

\renewcommand{\arraystretch}{1}

\clearpage

\clearpage

\clearpage

\clearpage

\clearpage

\clearpage

\clearpage

\clearpage

\clearpage

\clearpage

\end{document}